# Mass and Environment as Drivers of Galaxy Evolution II:
# The quenching of satellite galaxies as the origin of environmental effects

Ying-jie Peng[1], Simon J. Lilly[1], Alvio Renzini[2], Marcella Carollo[1]

1   Institute of Astronomy, ETH Zurich, 8093 Zurich, Switzerland
2   INAF - Osservatorio Astronomico di Padova, Vicolo dell'Osservatorio 5, 35122, Padova, Italy

ABSTRACT

We extend the phenomenological study of the evolving galaxy population of Peng et al (2010) to the central/satellite dichotomy in Yang et al. SDSS groups. We find that satellite galaxies are responsible for all the environmental effects in our earlier work. The fraction of centrals that are red does not depend on their environment but only on their stellar masses, whereas that of the satellites depends on both. We define a relative satellite quenching efficiency $\varepsilon_{sat}$, which is the fraction of blue centrals that are quenched upon becoming the satellite of another galaxy. This is shown to be independent of stellar mass, but to depend strongly on local overdensity, $\delta$, ranging between 0.2 and at least 0.8. The red fraction of satellites correlate much better with the local over-density $\delta$, a measure of location within the group, than with the richness of the group, i.e. dark matter halo mass.  This, and the fact that satellite quenching depends on local density and not on either the stellar mass of the galaxy or the dark matter halo mass gives clues as to the nature of the satellite-quenching process. We furthermore show that the action of mass-quenching on satellite galaxies is also independent of the dark matter mass of the parent halo. We then apply the Peng et al (2010) approach to predict the mass functions of central and satellite galaxies, split into passive and active galaxies, and show that these match very well the observed mass functions from SDSS, further strengthening the validity of this phenomenological approach. We highlight the fact that the observed *M** is exactly the same for the star-forming centrals and satellites and the observed M* for the star-forming satellites is independent of halo mass above $10^{12} M_\odot$, which emphasizes the universality of the mass-quenching process that we identified in Peng et al (2010). Post-quenching merging modifies the mass function of the central galaxies but can increase the mass of typical centrals by only about 25%.

## 1. INTRODUCTION

It has long been established that the properties of galaxies, to a greater or lesser degree, reflect their environments. As an example, the fraction of galaxies that are on the red sequence, $f_{red}$, is clearly a function of both stellar mass and environment (e.g., Kauffmann et al. 2003 & 2004; Baldry et al. 2006). It also correlates with other factors, such as surface mass density, as emphasized by Kauffmann et al (2003), although the direction of any causal link between quenching and mass density is unclear.

In a recent paper, Peng et al. (2010, hereafter P10), we developed a new, entirely phenomenological, approach to the study of the evolving galaxy population. This focused on the relative roles of stellar mass, $m$, and environment, $\rho$, in controlling the star-formation rates in galaxies. We especially considered the processes by which some galaxies apparently cease their star-formation almost entirely, a process we refer to as "quenching". Quenching should be distinguished from the gradual evolution of star-formation rates with cosmic time that is seen for all star-forming galaxies. Quenching is manifested in the fraction of galaxies that have the red colors characteristic of a passively evolving stellar population, $f_{red}$. Empirically, this is observed to be a strong function of both mass and environment.

Our phenomenological approach is based on identifying a limited number of underlying "simplicities" in the galaxy population, as revealed in the new generation of large surveys, both locally (e.g. SDSS, York et al 2000) and at significant look-back times (such as COSMOS/zCOSMOS, Scoville et al 2007, Lilly et al 2007, 2009). We then develop the analytic consequences of the underlying relations via simple continuity equations in order to identify the most basic features of galaxy evolution that are, in this sense, *demanded* by the data. The approach, by construction, contains almost no physics, but aims to identify the gross behaviors that any more physically-based model must satisfy and to give clues as to the principal physical processes involved.

An important part of the P10 analysis was the demonstration that the differential effects of galactic stellar mass and environment on $f_{red}$ are completely separable, at least to $z \sim 1$. This suggested that two distinct processes are operating, one we called "mass quenching", which is independent of environment, and one we called "environment quenching", which must be independent of stellar mass.

The empirical fact that the characteristic stellar mass of star-forming galaxies (i.e. $M^*$ of the Schechter function) does not change significantly since $z \sim 2$ (e.g. Bell et al. 2007; Pozzetti et al. 2010; Ilbert et al. 2010), despite the large, two order of magnitude, increase in the stellar masses of star-forming galaxies that are *not* quenched, requires that the mass-quenching must have a particular form. The mass-quenching rate, $\eta$, which defines the probability that a given star-forming galaxy is quenched in unit time, has to be proportional to the star-formation rate: $\eta = \mu \times SFR$. An entirely equivalent statement is that the survival probability of a given galaxy to grow to a certain stellar mass has to be $P(m) = \exp(-\mu m)$. The constant $\mu$ has a value $M^{*-1}$.

The link between the mass-quenching rate and the star-formation rate of galaxies implies that most mass-quenching occurs at high redshifts. At a given stellar mass, the quenching rate will be proportional to the sSFR, which is a factor of twenty or so higher at $z \sim 2$ than at the present epoch.

In P10, we defined an environment quenching efficiency $\varepsilon_\rho(m,\rho)$ that describes, at fixed stellar mass $m$, the increase of the red fraction as one moves to richer environments. We showed in P10 that this is empirically a function only of the environment variable and not of the stellar mass, $m$. We furthermore showed in P10 that if the over-density normalized to the average density of the Universe, $\delta$, is used as the environment measure, then $\varepsilon_\rho(\delta)$ is observed to be independent of epoch, at least to $z \sim 1$. The environmental differentiation within the galaxy population(s) nevertheless becomes stronger with time, since individual galaxies migrate to higher over-densities and the population occupies a broader range of densities as large scale structure develops in the Universe. Indeed, the *rate* of environment-quenching is given (to $z = 1$) by an epoch-independent function of density multiplied by the rate at which a given galaxy migrates to higher over-densities as large scale structure develops in the Universe.

We also suggested that a third quenching channel existed, namely "merger-quenching", associated with mergers of galaxies. In the P10 model, the rate of merging was assumed to be independent of stellar mass and the rate was taken from observational estimates of the merger rate of galaxies. The role of merging in the P10 model will be explored further in a future paper.

Our new phenomenological approach already had a number of encouraging successes in P10. These included offering beautiful "explanations" for the precise Schechter forms of the mass-functions of the star-forming and passive components of the galaxy population, in both high and low density regions, and providing remarkably precise predictions for the inter-relationships between the Schechter parameters that describe these various mass functions. Not least, the double-Schechter form for the overall mass function of galaxies (Baldry et al 2008) emerges naturally from the model.

This phenomenological model of the evolving galaxy population also offers natural explanations for the "anti-hierarchical" mass-age relation for passive galaxies, the run of α-enrichment with stellar mass, and offers a number of other predictions that can be compared with future observations. Of course, other approaches are also able to reproduce some or all of these phenomena, e.g. De Lucia & Blaizot (2007).

The environmental measure used in P10 was based on the distance to the 5th nearest neighbor above a given limiting luminosity, $M_{B,AB} < -19.3 - z$, enabling a self-consistent comparison of environments in SDSS locally and in zCOSMOS out to $z \sim 1$ (Kovač et al 2010). This adaptive density estimator typically samples the galaxy field on scales of about 1 Mpc and can be conveniently expressed as an over-density $\delta$ relative to the average number density of the tracer galaxies.

In the spirit of our phenomenological approach, we looked in P10 for possible physical processes that could account for the stellar-mass independence and epoch-independence of our environmental quenching effect. We pointed out in that paper that in the COSMOS mock catalogs of Kitzbichler & White (2007), which are based on semi-analytic modeling of the galaxy population within the Millenium cosmological dark matter simulation, the fraction of galaxies that are *satellites* of other galaxies, $f_{sat}(m,\delta,t)$ is also strongly dependent on environment, but, at fixed $\delta$, is almost independent of epoch (at least up to $z \sim 1$) and stellar mass. We therefore speculated on this basis that our "environment-quenching" might well be associated with the quenching of satellite galaxies and derived a

prediction for the δ-dependence of the fraction of satellites that are quenched, $f_{sat,red}(\delta)$

We define here "central" galaxies to be those galaxies that are the most massive and luminous galaxies within their dark matter haloes, irrespective of their spatial locations within their respective haloes. Other galaxies lying within the same dark matter halo are defined to be "satellites", again irrespective of any implied orbital parameters. To distinguish between centrals and satellites, all galaxies must therefore be assigned to a set of dark matter haloes, which are identified using a "group-finding" algorithm applied to the galaxy catalogue. In this paper we will refer to all such haloes as "groups" irrespective of the number of galaxies observed within them.

A distinction between central and satellite galaxies appears in many theoretical models for the evolution of galaxies. The *central* galaxies are expected to reside, more or less at rest, in the centers of the gravitational potential wells. When a smaller halo is accreted by a larger halo, it will become a "sub-halo", and the galaxy or galaxies in the smaller halo will all become *satellites* of the central galaxy in the larger halo. Some of these satellites may subsequently merge with the central. A central need not have any satellite galaxies, but almost all will, provided that one searches faint enough. It may be counter-intuitive to some readers that most galaxies are in fact centrals, regardless of their stellar masses, i.e. for every galaxy like the LMC that is a satellite of a Milky Way, there are more that dominate their own haloes. This can be seen later in this paper on Figure 13.

Many of the physical mechanisms that have been proposed to impose environmental effects on galaxies may operate primarily on satellite galaxies. These include strangulation, in which the fuel of a galaxy is removed through heating or stripping (Larson, Tinsley & Caldwell 1980; Balogh, Navarro & Morris 2000; Balogh & Morris 2000, Feldmann et al 2010), ram-pressure stripping, in which the interstellar medium itself is stripped away by motion through a high pressure intergalactic medium (e.g., Gunn & Gott 1972; Abadi, Moore, & Bower 1999; Quilis, Moore & Bower 2000), tidal stripping, and harassment, in which multiple high velocity encounters disrupt a galaxy (Farouki & Shapiro 1981; Moore et al. 1996).

Many recent observational analyses have therefore distinguished between satellites and centrals (e.g. Weinmann et al. 2006 & 2009; von der Linden et al. 2007; van den Bosch et al. 2008a & 2008b; Yang et al. 2009a & 2009b; Pasquali et al. 2009 & 2010; Skibba 2009; Hansen et al. 2009; Kimm et al. 2009). For example, van den Bosch et al. (2008a) showed that satellites are on average redder than centrals at the same stellar mass in the SDSS DR4 survey and concluded that on average 40% of star-forming satellites had been quenched.

We inferred in P10 that, if our (stellar mass independent) environment-quenching was in fact a phenomenon involving just the satellites, then the fraction of satellites that were quenched in this way would have to depend on our over-density parameter δ. We inferred a variation between about 20% (in the low density D1 quartile) to 80% (in the high density D4 quartile), averaging to 40%, as seen by van den Bosch et al (2008a). We were however unable to determine whether this variation with δ was direct or whether it simply reflected some correlation with other environmental parameters, such as group richness and/or dark matter halo mass.

With these motivations in mind, the current paper extends the P10 approach to satellites and centrals in the SDSS with the aims of:

(1) determining the underlying relationships, if any, between the environment-quenching of P10 and the satellite/central dichotomy by applying the P10 formalism of quenching "efficiencies" to satellites;

(2) identifying which environmental parameters appear to control the satellite quenching process;

(3) predicting the four mass functions of the centrals and satellites, split into passive and star-forming galaxies, as a test of this formalism.

The layout of the paper is as follows. In Section 2, at the request of the referee, we offer an extensive reprise of P10 that may be skipped by readers familiar with that paper. In Section 3, we summarize the basic input data that we have used. This is based on the same SDSS DR7 sample that we used in P10 with the added information from an SDSS DR7 group catalogue that has kindly been made available to us by Xiaohu Yang and collaborators. In Section 4, we examine the red fraction (at a given stellar mass), as a function of environment for centrals and satellites respectively, and show directly that the environment-quenching identified in P10 is indeed confined to satellite galaxies, as suspected. We then construct a "satellite quenching efficiency" that is directly analogous to the quenching efficiencies introduced in P10, and show that this is completely independent of stellar mass. In Section 5, we investigate the relations between the over-density δ and group richness and determine that it is the former that appears to control the environment-/satellite-quenching process. We also examine whether there is any difference in the mass-quenching process for satellites and centrals, concluding that there is none. In Section 6, we construct predictions for the mass functions of central and satellite galaxies and show that the inter-relationships between the Schechter parameters of these are in excellent agreement with the P10 predictions. This also derives a constraint on the amount of post-quenching merging that can have occurred in centrals and satellites.

Throughout the paper we use a concordance ΛCDM cosmology with $H_0 = 70$ km s$^{-1}$ Mpc$^{-1}$, $\Omega_\Lambda = 0.75$, and $\Omega_m = 0.25$. All magnitudes are quoted with the AB normalization, unless explicitly noted. We use the term "dex" to express the antilogarithm, i.e. 0.1 dex = $10^{0.1}$ = 1.259.

## 2. REPRISE OF P10

The P10 analysis built on the following three key observational facts about the galaxy population:
1. While the fraction of galaxies that are passive (red) is a strong function of both galactic mass $m$ and environment $\rho$, the differential effects of mass and environment on $f_{red}$ are fully separable. This empirical separability is seen by constructing two "relative quenching efficiencies": $\varepsilon_m(m,\rho)$ charts the effect of the environment at fixed mass and $\varepsilon_\rho(m,\rho)$ traces the effect of mass at fixed environment. By separability, we mean that $\varepsilon_m(m,\rho)$ is actually independent of environment and likewise that $\varepsilon_\rho(m,\rho)$ is independent of mass. This result was shown to hold both in the SDSS and zCOSMOS at least out to $z \sim 1$. In our P10 work, we implicitly assume that this separability is also maintained at still higher redshifts.
2. The shape of the mass function of star-forming galaxies is remarkably constant in terms of the Schechter M* and $\alpha_s$, while φ* increases with time. This has been clearly demonstrated to redshifts of at least $z \sim 2$. This requires a particular form for the quenching of galaxies.

3. The specific star formation rate sSFR($m, t$) of star-forming galaxies is at most a weak function of stellar mass and falls sharply between $z = 2$ and the present. The sSFR($m,t$) is also evidently independent of environment up to $z \sim 1$ for star-forming galaxies. The simple behavior of the sSFR with mass and environment greatly simplified our analysis, but is not strictly required for the validity of most of the conclusions.

The parameters used in P10 are summarized in Table 1.

*Table 1: Summary of parameters in P10*

| Parameter | Definition |
|---|---|
| $m$ | galaxy stellar mass |
| $\rho$ | generic galaxy *external* environment. In P10 and in the current paper, we choose to evaluate the environment $\rho$ as the over-density $\delta$. We then use $\rho$ and $\delta$ interchangeably. $\rho$ can also be computed with other environment estimators, such as the distance to the group center, group halo mass etc. |
| SFR($m,t$) | star formation rate |
| sSFR($m,t$) | specific star formation rate |
| $\beta$ | the logarithmic slope of the sSFR–mass relation: sSFR $\sim m^\beta$ |
| M* | the characteristic mass of the Schechter function |
| $\alpha_s$ | the faint end slope of the Schechter function |
| $\alpha(m)$ | the logarithmic slopes of the mass function: $\alpha = (1+\alpha_s) - m/M^*$ for a Schechter function |
| $\varepsilon_\rho(\rho,t)$ | the relative environmental quenching efficiency |
| $\varepsilon_m(m,t)$ | the relative mass quenching efficiency |
| $\lambda_m(m,t)$ | mass-quenching rate: $\lambda_m(m,t) = \mu \text{SFR}(m,t) = \text{SFR}(m,t)/M^*$, which is the probability that a given galaxy is quenched per unit time. |
| $\mu$ | a constant, presumably reflecting the physical process of mass-quenching, which produces a value of $M^* = \mu^{-1}$ |
| $\lambda_\rho(\rho, t)$ | environment-quenching rate |
| $\kappa_+(\rho, t)$ | major merger rate - the merging influx into a given mass bin, normalised to the number of galaxies in that bin |
| $\kappa_-(\rho, t)$ | major merger rate - the merging outflux out of a given mass bin, normalised to the number of galaxies in that bin |
| $\eta_\rho(\rho, t)$ | combined mass-independent quenching rate: $\eta_\rho = \lambda_\rho + \kappa_-$ |
| $\eta_m(m, t)$ | combined environment-independent quenching rate: $\eta_m = \lambda_m$ |
| $\eta(m,\rho, t)$ | combined quenching rate: $\eta = \eta_m + \eta_\rho = \lambda_m + \lambda_\rho + \kappa_-$ |

Note: for parameters contain $\rho$ such as $\varepsilon_\rho(\rho)$, the $\rho$ in the subscript represents the generic property of the parameter, which is directly related to the generic environment $\rho$. While the $\rho$ in the parentheses can be computed with different environment estimators. For example in P10 and in the current paper, we choose to evaluate the environment $\rho$ as the over-density $\delta$, we then write $\varepsilon_\rho(\delta)$ as $\varepsilon_\rho(\rho)$ and we use them interchangeably.

The separability of $m$ and $\rho$ in determining $f_{red}$ in SDSS and surveys at higher redshift is most naturally understood if two distinct processes are occurring, which we refer to as "mass-quenching" and "environment- quenching". Regardless of the physical origin of these two processes, we can assign a quenching rate to each of them, which is simply the probability that a given star-forming galaxy will cease star-formation per unit time. In P10, we used the notation $\lambda_m$ and $\lambda_\rho$ to indicate these quenching rates.

In P10, we also considered a merging channel associated with quenching following a major merger. This was also characterized by a rate, $\kappa$. In P10, we differentiated between the influx into a given mass bin, $\kappa+$ and the outflux out of the bin, $\kappa-$, which represented the destruction of galaxies (which we assumed was accompanied by quenching). Both $\kappa$ were expressed as a fractional rate relative to the number of galaxies in the bin. We assumed in P10 that these merging rates were independent of $m$. We will return to the whole question of merging in a future paper.

In P10, the combined quenching rate for a given galaxy, i.e. the sum of $\lambda_m$, $\lambda_\rho$ and $\kappa$- was called $\eta$. Needless to say, all these probabilities should be understood as reflecting the action of these processes on a population of galaxies.

The essence of P10 was to constrain the nature of the two processes, mass-quenching and environment-quenching, by applying the basic continuity equation(s) in stellar mass and environment subject to the constraint of the simple observational facts listed above. Separability means that we can look at mass-quenching and environment-quenching, independently.

Looking first at environment quenching. The fact that the environment quenching efficiency $\varepsilon_\rho(\rho)$, when evaluated at fixed over-density $\delta$, is the same at $z \sim 1$ in zCOSMOS as in SDSS locally, indicates that environment-quenching occurs phenomenologically as large scale structure develops in the Universe, i.e. as galaxies migrate to regions of higher over-density. The environment-quenching term for a given galaxy could therefore be written in terms of the migration of a galaxy towards regions of higher over-density, i.e. $d\rho/dt$.

Because the fraction of satellites *at the same fixed over-density* is also largely independent of mass and epoch, at least in the Millennium mock catalogues, we postulated in P10 that our environment-quenching process could be acting through satellite galaxies. This is the hypothesis that we test in the current paper.

Because mass-quenching is, from separability, the only quenching channel that depends on stellar mass, mass-quenching will control the shape of the mass-function $\phi(m)$ of the surviving star-forming galaxies. The clearest view of the action of mass-quenching in fact comes from analysis of $\phi(m)$ of the surviving star-forming galaxies.

To maintain the observed constancy of the shape of $\phi(m)$ requires (equation 10 of P10) a particular form of the mass-quenching rate $\lambda_m$ and $\kappa_-$ (the elimination of galaxies through merging). The mass dependence of these must exactly cancel the mass-dependence of a term that is given by $(\alpha+\beta)$sSFR, where $\alpha$ is the local logarithmic slope of the star-forming mass function, the mass-dependence of which is given by $m/M^*$ for a Schechter function. We have assumed in P10 that the $\kappa_-$ term is independent of stellar mass. Then, the assumptions (observational axioms) that the shape of the star-forming mass function, as defined in terms of $\alpha_s$ and $M^*$, is constant with time and that $\beta \sim 0$, then requires that $\lambda_m$ itself has the specific form that is proportional to $m$ times the sSFR, i.e. to the SFR.

$$\lambda_m(m,t) = \mu \text{SFR}(m,t) = \text{SFR}(m,t)/M^*. \qquad (1)$$

where $\mu$ is a constant, presumably reflecting the physical process of mass-quenching, which produces a value of $M^* = \mu^{-1}$. In our phenomenological approach the link between $\mu$ and $M^*$ appears the other way around: the value of $\mu$ is "set" by the observed constant value of $M^*$, but physically it is presumably the properties of mass-quenching which set $M^*$.

We stress that equation (1) above (i.e. equation (17) in P10) is the direct mathematical consequence of the observationally defined axioms (or assumptions) listed above. A detailed derivation from the continuity equations is given in the Appendix A.

If this mass-quenching formula in equation (1) holds at all masses, then it is easy to see that it produces a mass-function of newly mass-quenched galaxies that has exactly the same $M^*$ as that of the star-forming galaxies, but has a different faint-end slope $\alpha_s$, with $\Delta\alpha_s \sim 1$. Because $M^*$ and $\alpha_s$ for star-forming galaxies are observed to be invariant with time since early epochs, it follows that the integrated $\phi(m)$ of the mass-quenched passive population will also build up over time with the same constant $M^*$ and modified $\alpha_s$. In fact, the increase in log $\phi^*$ of the passive population over time exactly matches that of the star-forming galaxies (provided $\alpha_s < -1$) producing a characteristic double Schechter function shape that once established should not change (see the Appendix B in the current paper for a simple derivation of this fact).

Correspondingly, in low density environments, where the effects of environment-quenching are negligible, the close relationship between the two $\phi(m)$ for the active and passive populations that is produced by applying the above equation for $\lambda_m$ over all masses, produces a characteristic shape of $f_{red}(m)$ and thus $\varepsilon_m(m)$. This is due to the precisely defined difference in the faint-end slopes of the two Schechter functions. This characteristic shape for $\varepsilon_m(m)$ is precisely observed in SDSS, in all environments, indicating that the above formula for $\lambda_m$ must indeed hold over all masses, modulo the rather weak "assumption" discussed above.

Apart from the weak assumption about any mass-dependence of $\kappa_-$, the form of the mass-quenching law given by equation (1) is therefore not an assumption, but is required by the data, i.e. by the constancy of the shape of the star-forming mass-function and by the difference in $\alpha_s$ between active and passive galaxies, equivalent to $f_{red}$ in low density environments, more generally, to $\varepsilon_m(m)$.

Again, several commentators have assumed that the model would work equally well with any quenching law with a power-law mass-dependence $m^\beta$, i.e. that the strong epoch-dependence that is implied in equation (1) by the strong cosmic evolution of the star-formation rates is not required. This is incorrect. Such a quenching law would indeed produce, instantaneously, the correct relation between the Schechter $\phi(m)$ of star-forming galaxies and the $\phi(m)$ of the most recently formed passive galaxies. But, if the time-dependence of mass-quenching is not precisely matched to that of the star-formation rates, then the $M^*$ of the star-forming population will not be constant, as required, and the mass-function of the integrated passive population would have to be represented by a complicated integral over time.

In Section 7 of P10, we emphasized the formal equivalence of the mass-quenching rate defined as above in equation (1) in terms of a star-formation rate, and a quenching law that was in effect a limit to the stellar masses of galaxies. If we write such a limit in terms of a survival probability for a galaxy to reach a mass stellar mass $m$ without being mass-quenched, then it is easy to see that equation (1) produces

$$P(m) = e^{-\mu m} = e^{-m/M^*} \qquad (2)$$

Since stellar mass is closely correlated to the dark matter halo mass, one could also view this equivalence in terms of a limit to the dark matter halo mass that can support star-formation. This is a widely accepted idea, motivated by theoretical considerations, but not yet observationally established. In our phenomenological approach, equations (1) and (2) are equivalent descriptions of the mass-quenching process.

As discussed in P10, a physical mass-limit interpretation of mass-quenching, i.e. equation (2), while formally completely equivalent to equation (1), must nevertheless explain why the survival probability has this precise form over two or more orders of magnitude in stellar mass, and not some other. Furthermore, any second-parameter that is introduced to produce scatter into the limiting stellar mass should, from separability, be independent of environment. This latter argument traces back to the fact that $M^*$ for star-forming galaxies is observationally independent of environment as analyzed in P10 and as explored further later in the current paper.

The equivalence discussed in the previous two paragraphs is an example of how the approach in P10 can only be interpreted phenomenologically. The phenomenological linkage between the mass-quenching rate and the star-formation rate does not necessarily imply any physical or causal connection between these two. What can however be said, with some certainty, is that the actual quenching rate, which is a well-defined quantity, must (within a suitable tolerance) have exactly the same dependence on stellar mass and time as the SFR. If it doesn't, then the resulting population of galaxies will not exhibit the simple observational features that are the basis of the analysis.

The model in P10, which is based on a handful of simple observational inputs successfully reproduces many of the gross features of the galaxy population. In particular our simple empirically based model naturally produces:

1. A single Schechter mass function for star-forming galaxies with an exponential cutoff at a value of M* that is set uniquely by the constant of proportionality between the SFR and mass quenching rates. This value of M*, which is established solely by mass-quenching, should be independent of environment, and (by construction) independent of epoch over long periods of cosmic time.
2. A double Schechter function for passive galaxies with two components. The dominant component at high masses is produced by mass quenching and has exactly the same M* as the star-forming galaxies but a faint end slope that differs by $\Delta\alpha_s \sim 1$. The other component is produced by environment effects. These are independent of mass, so this component has both the same M* and the same $\alpha_s$ as the star-forming galaxies, but an amplitude that is strongly dependent on environment.
3. The total mass function, summing active and passive populations, is also inevitably a double Schechter function.

All of these detailed predictions for the inter-relationships between the Schechter parameters of the star-forming and passive galaxies, across a broad range of environments, are indeed seen to high accuracy in the SDSS, lending strong support to the validity of this approach, and indicating that our "model" should offer a good description of the most basic features of the evolving galaxy population.

This simple model also naturally reproduces several qualitative features of the galaxy population, including the "anti-hierarchical" age-mass relation for passive galaxies and the qualitative variation of formation timescale indicated by the relative α-element abundances.

This analytic framework enabled also in P10 to establish predictions for quantities such as the mass function of the population of any set of transitory objects that are in the process of being mass-quenched.

Although the model is purely phenomenological, it makes clear what the evolutionary characteristics of the relevant physical processes must follow and any more physically based model must obey.

## 3. OBSERVATIONAL DATA

### 3.1 SDSS Sample

The sample of galaxies analyzed in this paper is the same SDSS DR7 (Abazajian et al. 2009) sample that we constructed in P10. Briefly, it is a magnitude-selected sample of galaxies that have clean photometry and Petrosian SDSS *r*-band magnitudes in the range of $10.0 < r < 18.0$ after correcting for Galactic extinction. The parent photometric sample contains 1,579,314 objects after removing duplicates, of which 238,474 have reliable spectroscopic redshift measurements in the redshift range of $0.02 < z < 0.085$. Each galaxy is weighted by $1/\text{TSR} \times 1/V_{\max}$, where TSR is a spatial target sampling rate, determined using the fraction of objects that have spectra in the parent photometric sample within the minimum SDSS fiber spacing of 55 arcsec of a given object. The $V_{\max}$ values are derived from the *k*-correction program v4_1_4 (Blanton & Roweis 2007). The use of $V_{\max}$ weighting allows us to include representatives of the galaxy population down to a stellar mass of about $10^9 M_\odot$.

Rest-frame absolute magnitudes for the SDSS sample are derived from the five SDSS *ugriz* bands using the *k*-correction program (Blanton & Roweis 2007). All SDSS magnitudes are further transformed onto the AB magnitude system. The stellar masses are determined directly from the same *k*-correction code with Bruzual & Charlot (2003) population synthesis models and a Chabrier IMF.

The SFRs of the SDSS blue star-forming galaxies were derived by Brinchmann et al. (2004, hereafter B04). These are based on the Hα emission line luminosities, corrected for extinction using the Hα/Hβ ratio, and corrected for aperture effects. The B04 SFRs were computed for a Kroupa IMF and so we convert these to a Chabrier IMF, by using log SFR (Chabrier) = log SFR (Kroupa) − 0.04.

### 3.2 Density estimates

As in P10, one approach to characterizing the environment of a given galaxy is by a dimensionless density contrast $\delta_i = (\rho_i - \rho_m)/\rho_m$. Here $\rho_i$ is an estimate of the local density around the $i^{th}$ galaxy and $\rho_m$ is the mean density at that redshift. The densities $\rho_i$ are computed from the volume of that cylinder, centered on each galaxy with length ±1000 km s$^{-1}$ and with an adjustable radius, which contains the five closest neighbor galaxies with $M_{B,AB} \leqslant -19.3 - z$. As discussed in P10, one advantage of this choice of tracer galaxies is to make our SDSS density field directly comparable to the zCOSMOS density field up to z~0.7 by trying to include the same galaxies at every redshift (see Kovac et al. 2010 for a full discussion). Since zCOSMOS is I-band selected it makes sense to use a rest-frame B-band luminosity. We use −z as a luminosity modifier to approximately account for the luminosity evolution of both passive and active galaxies. Empirically, this leads to a more or less constant mean comoving density of tracer galaxies, i.e. a constant $\rho_m$ over cosmic time. Thus in our analysis the over-density $\delta$ is equivalent to the actual comoving density $\rho$, with the same relationship at all epochs and we will use these interchangeably. This choice of tracer galaxies is also complete for all galaxy colors throughout the $0.02 < z < 0.085$ range, avoiding any $V_{max}$-type correction to the density estimates. We choose a "unity-weighted" density without any mass or luminosity weighting of the galaxies within the sampling cylinder.

## 3.3 Group Catalog

The group catalogue that we use in this work is an SDSS DR7 group catalogue kindly made available by Yang et al. This is the updated version of the Yang et al. SDSS DR4 group catalogue, described in Yang et al. (2005, 2007, hereafter Y07). It applies a rather sophisticated iterative algorithm, calibrated on mock catalogues (see Yang et al. 2005), to the NYU Value-Added Galaxy Catalog (NYU-VAGC, Blanton et al. 2005), which is also based on the SDSS DR7 sample. Y07 selected all galaxies in the NYU-VAGC Main Galaxy Sample with redshifts in the range of $0.01 \leqslant z \leqslant 0.20$ and with a redshift completeness greater than 0.7, i.e. a broader range of redshift than we will use.

Three group samples are constructed from these galaxies. Sample I contains 599,451 galaxies with SDSS redshifts only. Sample II contains 602,729 galaxies and is composed of Sample I plus additional redshifts from alternative surveys like 2dFGRS (Colless et al 2001) etc. Sample III contains 639,555 galaxies and is composed of Sample II plus additional galaxies that lack redshifts due to fiber collisions and are assigned redshifts of their nearest neighbors. For each of these three galaxy samples, two group catalogues are constructed based respectively on the "Petrosian" and "Model" absolute magnitudes of the galaxies from the NYU-VAGC. There are thus available six different group catalogues in total. The results presented in this work are based on the Sample II with the "model" magnitudes. We also repeated our analysis with other group samples and found completely consistent results and only insignificant changes and so this particular choice is rather arbitrary.

All galaxies can then be classified as either "central" galaxies or "satellite" galaxies. The centrals include single galaxies that do not have identified companions above the SDSS flux limit. In order to reduce the contamination of the central sample by spurious interlopers into the group, we also required that the "central" galaxies be simultaneously both the most massive and the most luminous (in the R-band) galaxy within a given group. Where multiple objects lie in the same "group", all other members of the group are then identified as satellites. All singletons are automatically centrals. This operational definition of central eliminates a small fraction (2.1%) of galaxies that would have been classified as centrals using only a mass or luminosity criterion on its own. Including these ambiguous galaxies in the set of centrals also produces indistinguishable changes to the results presented in this paper, and is thus not of great importance. It also means that some groups actually have no "central" galaxy, and consist only of satellites. Despite the nomenclature of "central" galaxy, it should be noted that the spatial position of a galaxy is not used in the classification of centrals and satellites (see Skibba 2011).

No group-finder that aims to identify dark matter haloes from the locations of galaxies in $(ra,dec,z)$ space can be expected to operate perfectly, even when calibrated against mock catalogues in which the underlying dark matter distribution is known. It should be appreciated that any "over-fragmentation", or "over-merging", of groups will lead to mis-classification of satellites as centrals, and vice versa. This may produce a bias reducing the observed differences between centrals and satellites. More detailed discussion about the completeness and contamination of the satellite and central can be found in Weinmann et al (2009). Some second-order effects may introduce a redshift-dependence on the performance of the group-finder, e.g. due to the changing number density of galaxies, or to issues associated with the spatial scale of "fibre-collisions" (see the discussion in Yang et al. 2007). A full consideration of these effects is beyond the scope of the paper.

It is important to appreciate, given the above definitions, that the centrals and satellites will have different distributions of halo mass, and of other environment measures. This is because the centrals of those haloes that contain the observed satellites constitute only a small fraction of all centrals since the latter include also the "singletons" whose satellites are (presumably) too faint to be observed. Not least, at a given stellar mass, the satellites will generally be in more massive haloes than the centrals.

The flux limit of SDSS requires some consideration. The binary classification of the SDSS galaxies into centrals or satellites should remain valid down to the SDSS flux limit, and should not depend on redshift, since for any galaxy the classification as a satellite rests only on the existence or otherwise of brighter (or more massive) galaxies, and not on fainter ones. In constructing mass functions for the set of centrals and satellites we must of course correct for the varying mass completeness of the sample with redshift, but we can use the standard $V_{max}$ approach for this. In constructing color fractions, e.g. the fraction of red galaxies, $f_{red}$, at different stellar masses and in different environments, it is clearly safest to only consider those galaxies that lie above the mass completeness limit at their redshift that is given by the masses of the reddest galaxies (assumed to have the highest M/L ratio). These then enter the calculation of the red fraction at their stellar mass without weighting.

The group catalogues are then cross-matched with our P10 galaxy sample, which was constructed with more strict selection criteria. Construction of the other physical properties of the galaxies such as stellar mass, absolute magnitudes, color, over-density is done in a way that is completely consistent with the P10 analysis.

In the present work, we define the Richness $R$ of a given group to be that number of spectroscopically confirmed members that lie above the same luminosity limit as used in the definition of the density field, i.e. $M_{B,AB} \leqslant -19.3 - z$. This choice makes it relatively straightforward to compare the two measures of environment, as discussed further in Section 5. It also means that the Richness is redshift independent over $0.02 < z < 0.085$ and does not involve $V_{max}$ corrections. This choice does however mean that many of the Yang et al groups with multiple members have Richness = 0.

Towards the end of the paper we will also examine halo mass. We use for this purpose the dark matter halo mass $M_h$ estimated by Yang et al from the characteristic integrated luminosity or the characteristic stellar mass of the group (see Yang et al 2007 for details) calibrated against the dark matter masses of groups found in mock catalogues. The characteristic luminosity/stellar mass is defined as the combined luminosity/stellar mass of all group members above some luminosity limit and is further corrected for the completeness of the survey. Since the characteristic stellar mass is less affected by the ongoing SFR of the galaxy, it is expected to be a better halo mass indicator than the characteristic luminosity, which is shown in the Fig. 5 of Yang et al (2007). The results presented in this paper are based on the halo masses estimated from the characteristic stellar mass of the groups.

# 4. THE RED FRACTION AND THE CENTRAL - SATELLITE DICHOTOMY

If the environmental effects in the galaxy population that we identified and isolated in P10 are indeed due to the central-satellite dichotomy, i.e. if all environmental effects act only on satellites, then we would expect to see several things.

First, since the centrals can *only* be mass-quenched, the red fraction of centrals should be independent of their environments, but a strong function of stellar mass. Specifically, the red fraction as a function of mass will be given solely by the mass-quenching efficiency $\varepsilon_m(m)$ of P10.

Second, since satellites can be quenched *either* by mass- or environment-quenching, the red fraction of satellites will depend on both mass and environment. However, analogously to our analysis in P10, the *differential* effects of the environment on the satellites, i.e. normalized relative to the properties of the centrals at the same mass, should be independent of stellar mass.

In this section, we test these expectations by applying a similar methodology to that in P10 to the SDSS sample of galaxies split, as described in Section 3, into centrals and satellites. The galaxy population is further divided, exactly as in P10, into "blue star forming" and "red passive" galaxies based on their rest-frame $(U - B)$ colors, dividing at a threshold color that is given by equation (2) in P10.

In this section we will first investigate the dependence of the galaxy red fraction on the central-satellite dichotomy with the SDSS DR7 Yang et al. group catalogue. Then we introduce the satellite quenching efficiency in section 4.2. In section 4.3 we derive a new analytical form of the mass-quenching efficiency that is useful later in the paper.

## 4.1 Red Fraction of Centrals and Satellites

Fig. 1 shows the red fraction of the centrals and of the satellites as a function of over-density $\delta$ for different stellar masses. The red fractions are determined using a sliding box of width 0.3 dex in mass and 0.3 dex in $(1+\delta)$, these intervals being chosen to be comparable to the uncertainties in mass and over-density. The hatched regions around selected lines show typical observational sampling uncertainties simply derived from the binomial error of the fraction at a 68% confidence level.

As discussed above, the sample of centrals occupies a different range of over-density $\delta$ than the sample of satellites. Only 10% of centrals have $\log(1+\delta) > 1.0$, which is about the median of the $\delta$ distribution for the satellites.

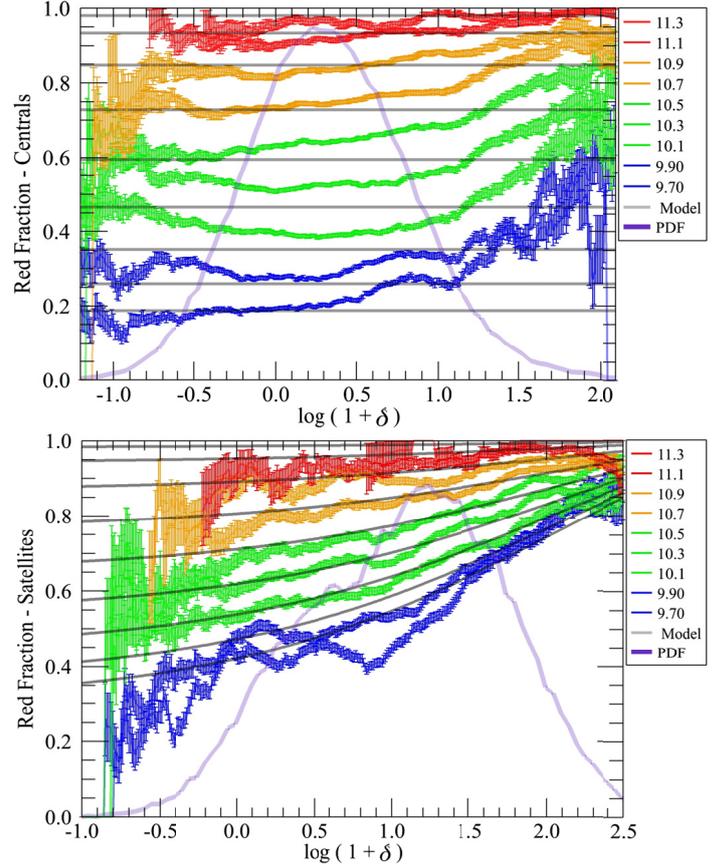

Figure 1. The red fraction as a function of over-density $\delta$ at different galaxy masses for central galaxies (top) and for satellite galaxies (bottom). The light purple curves in both panels show the overall distribution of centrals and satellites in over-density. The red fractions of centrals are independent of over-density over the range of $-1.0 < \log(1+\delta) < +1.0$, where 90% of the central galaxies are found - the deviations at higher densities may reflect limitations in the group catalogue (see text). In contrast, the red fractions of the satellite galaxies steadily increase with both increasing density and with increasing mass. The grey lines in each diagram correspond to the simple parameterizations given in the text in equation (6).

It is evident in Fig. 1 that the red fraction of the centrals, at a given stellar mass, does not change significantly with environment over the density range of $-1.0 < \log(1+\delta) < 1.0$ that contains 90% of the central galaxies. In the centrals occupying the highest 10% over-densities, i.e. those with $\log(1+\delta) > 1.0$, the red fraction does apparently increase, especially for lower mass galaxies. We strongly suspect this could reflect difficulties in the identification of true centrals due to over-fragmentation of groups by the group finding algorithm, which would lead to some satellites being misidentified as centrals. This would in turn tend to increase the red fraction of centrals since satellites are on average redder. This effect would be expected to be most severe for low mass galaxies in high-density regions. Indeed, we find that almost all of the low mass $m < 10^{10}$ M$_\odot$ "centrals" with $\log(1+\delta) > 1.5$ lie within 500 kpc of a more massive galaxy.

It should also be recognized that the physical transformation of a central into a satellite as a halo merges with a larger one may not exactly follow the simple operational definitions that we of necessity use and so low-mass centrals on the outskirts of larger haloes may well already be affected by the larger halo. Given this concern, we therefore choose to ignore this regime of mass and density.

The horizontal solid lines in the left panel of Fig. 1 show the mass-quenching efficiency $\varepsilon_m$ from P10, as determined from the *entire* galaxy population. As discussed in P10, $\varepsilon_m$ is a reflection of the mass-quenching process, and gives the red fraction of the *general* population in the most *under*-dense regions, i.e. log $(1+\delta) < 0$. Fig. 1 confirms that $\varepsilon_m$ also represents the red fraction of *centrals* over a much broader range of δ, i.e. at least to log $(1+\delta) \sim +1.0$, as expected if these central galaxies are *only* mass-quenched, even in the highest density environments.

By comparison, the bottom panel of Fig. 1 shows the red fraction of satellites as a function of environment for different stellar masses. It is clear that the red fraction monotonically increases with increasing mass and density, which is expected as satellite galaxies may have been quenched through either mass-quenching or environment-quenching.

### 4.2 Satellite quenching efficiency

As in P10, it is illuminating to look at the *differential* effects of environment on the red fraction of the satellite galaxies by normalizing to a blue fraction at the same stellar mass. To do this, we define the *relative satellite quenching efficiency* $\varepsilon_{sat}$ in a similar way to the construction of the relative environment- and mass- quenching efficiencies, $\varepsilon_\rho$ and $\varepsilon_m$, in P10.

The $\varepsilon_{sat}$ is an observationally defined quantity that could, in principle, be a function of both stellar mass and environment. To obtain $\varepsilon_{sat}$, we look, at a given stellar mass $m$, at the fraction of star-forming satellites compared to the fraction of star forming *central* galaxies. In P10, the environment-quenching efficiency was normalized to the blue fraction in the most under-dense regions. Here we normalize to the blue fraction of the centrals, which we assume (from Fig. 1) to be independent of environment, ρ, and to be given by $\varepsilon_m(m)$. With this definition, $\varepsilon_{sat}$ gives the fraction of (previously central) star-forming galaxies that are quenched because they become satellites of another galaxy.

$$\varepsilon_{sat}(\rho,m) = \frac{f_{sat,red}(\rho,m) - f_{cen,red}(m)}{f_{cen,blue}(m)}$$
$$= \frac{f_{sat,red}(\rho,m) - \varepsilon_m}{(1-\varepsilon_m)} \quad (3)$$

Fig. 2 shows the observed values of $\varepsilon_{sat}$ computed using the second equation, combining the observational data for the satellites from Fig. 1 with the curve of $\varepsilon_m(m)$ taken from P10. Clearly these curves largely overlap, indicating that $\varepsilon_{sat}$ is independent of stellar mass. A fit to the observed $\varepsilon_{sat}$ averaging over all stellar masses is shown as the thick black line in Fig. 2.

The independence of $\varepsilon_{sat}$ on stellar mass is shown more directly in Fig. 3 where we show the global $f_{red}$ for centrals and satellites (on the left-hand axis), obtained by summing over all environments, and (on the right-hand axis) the environment-averaged satellite quenching efficiency defined as

$$\overline{\varepsilon}_{sat}(m) = \frac{\overline{f}_{sat,red}(m) - \overline{f}_{cen,red}(m)}{\overline{f}_{cen,blue}(m)} \quad (4)$$

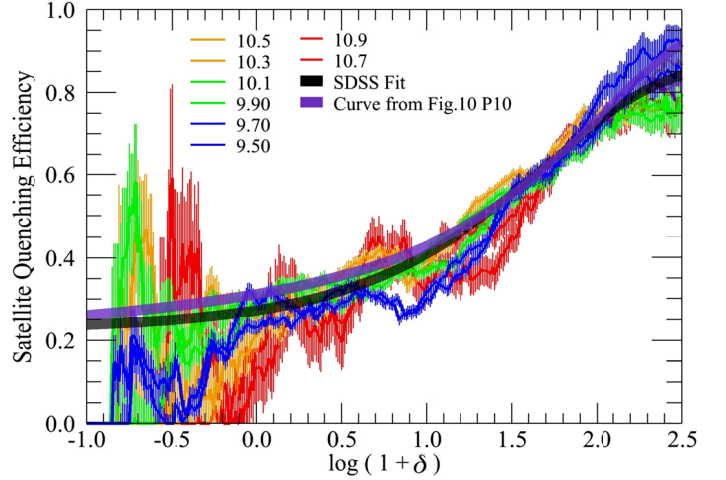

Figure 2. Observed values of the relative satellite-quenching efficiency, $\varepsilon_{sat}(\rho)$ for different galaxy masses. These curves are all essentially identical and show that the satellite-quenching is operating independently of stellar mass. The black thick curve is fitted satellite quenching efficiency for all stellar masses. The purple curve is taken from Figure 10 in Peng et al. 2010, which shows the predicted fraction of quenched satellites from that paper.

This quantity is evidently constant over a wide range of stellar mass, deviating only when the noise becomes very large at high masses, when almost all galaxies are red. The remarkable constancy of the average satellite quenching efficiency with stellar mass (averaged over all environments) strongly supports the impression from Fig. 2. The average value is around 40%, which agrees with the global analysis of van den Bosch et al. (2008), who did not include environment as a variable.

As noted above, the quantity $\varepsilon_{sat}$ is the fraction of previously blue centrals that are quenched on becoming a satellite. Of course, this could, in our approach, be a one time "throw of the dice" when a given galaxy becomes a satellite, or it could represent a temporal "duty cycle" in which a given galaxy spends part of its life active, and part passive.

It was argued in P10, that if our environmental quenching is entirely due to processes that act on satellite galaxies, then there will be a simple linkage between $\varepsilon_\rho$, $\varepsilon_{sat}$ and $f_{sat}$, the fraction of galaxies that are satellites. In P10, we showed that both $\varepsilon_\rho$ and $f_{sat}$ are, in the data and in Millenium mock catalogues respectively, independent of stellar mass and epoch, at least to $z \sim 1$ (see Fig. 5 and Fig. 10 in P10), implying that $\varepsilon_{sat}$ should also be independent of stellar mass, as seen in Figs. 2 and 3, and also of redshift, which can be tested in the future. In other words,

$$\varepsilon_{sat}(\rho) = \frac{\varepsilon_\rho(\rho)}{f_{sat}(\rho)} \quad (5)$$

The "predicted" $\varepsilon_{sat}$ that was derived on Fig. 10 of P10 using equation (5), is shown as the thick purple line in Fig. 2.

This shows excellent agreement with the empirically fitted black line.

The conclusion from the above is that the dominant environmental effects that we identified in P10 are indeed *entirely* driven by the quenching of satellite galaxies. Furthermore, the fraction of satellites that are quenched, relative to the number of star-forming central galaxies, is independent of stellar mass but does depend quite strongly on the over-density as parameterized by our δ parameter. We show below that δ most likely reflects the location within a halo rather than the mass of a halo.

Rearranging equation (3), the red fraction of the satellites is given by

$$f_{sat,red}(\rho,m) = [1-f_{cen,red}(m)]\varepsilon_{sat}(\rho) + f_{cen,red}(m)$$
$$= (1-\varepsilon_m(m))\varepsilon_{sat}(\rho) + \varepsilon_m(m)$$
$$= \varepsilon_m + \varepsilon_{sat} - \varepsilon_m\varepsilon_{sat} \quad (6)$$

whereas the red fraction of the centrals will be given by the simple mass-quenching efficiency. We can use the fitted $\varepsilon_{sat}$ from the data (black line in Fig. 2), with $\varepsilon_m$ determined from P10, to determine the overall red fractions of centrals and satellites as functions of mass and environment via equation (6). The results are shown in the two panels of Fig. 1 as the grey lines labeled as "model". Again, excellent agreements are seen. It is worth noting in passing that since $\varepsilon_{sat}$ must lie in the range of zero to one, a requirement is that $f_{sat} \geq \varepsilon_\rho$ in all environments. Inspection of Fig. 10 in P10 shows that indeed this appears to be the case.

We would also expect the function $\varepsilon_{sat}(\rho)$ to be independent of epoch. However, this does not imply that the effects of satellite-quenching on the galaxy population are unchanging with cosmic time. The number of galaxies that are satellites will monotonically increase due to the hierarchical assembly of haloes. Equivalently, even though $f_{sat}(\rho)$ should be largely redshift-independent, the median over-density of galaxies is expected to increase with cosmic time, thereby increasing $f_{sat}$, and possibly increasing the mean δ of the satellites. For instance, in Figure 2, a mean satellite quenching efficiency <$\varepsilon_{sat}$> ~ 40% corresponds to an over-density log(1+δ) ~ 1.0, which is indeed the mean over-density of satellite galaxies in our SDSS sample. If the median over-density of satellite galaxy population changes with cosmic time and if $\varepsilon_{sat}$ is independent of epoch as well as mass, then the <$\varepsilon_{sat}$> should also change, producing a corresponding change in the mean red fraction of satellites <$f_{red,sat}$>. This will be examined using a new zCOSMOS group catalogue in a later paper.

### 4.3 Analytic expression for the mass quenching efficiency $\varepsilon_m$, the red fraction of centrals

The relative mass- and environment- quenching efficiencies, $\varepsilon_m$ and $\varepsilon_\rho$, were introduced in P10 as empirical quantities, as is $\varepsilon_{sat}$ in this paper. In this section we derive an analytic expression for the expected mass-quenching efficiency function $\varepsilon_m$, which is the red fraction of central galaxies, and also the red fraction of all galaxies in very low density environments. This is based on the simple form of the mass-quenching process proposed in P10, and will be valid for the simple case of β = 0, where β is the index of the sSFR-mass relation, $sSFR \propto m^\beta$.

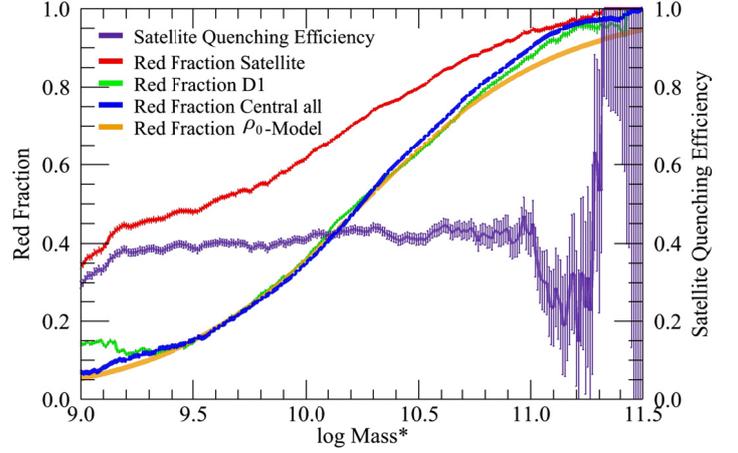

Figure 3. The red fraction as a function of stellar mass for satellites (red line), centrals (blue line) and (from P10) for all galaxies in the low-density D1 quartile (green line). The red fraction in the lowest density environment ($\rho_0$) determined from equation (8) is plotted as the dark orange line labeled as "$\rho_0$-Model". The satellite quenching efficiency determined from equation (3) (purple line) is largely independent of stellar mass.

From the continuity equation (10) in P10, the change of the red fraction at fixed mass and environment is given by

$$\left.\frac{df_{red}}{dt}\right|_{m,\rho} = \frac{1}{N_{tot}}\left.\frac{dN_{red}}{dt}\right|_{m,\rho} - \frac{N_{red}}{N_{tot}^2}\left.\frac{dN_{tot}}{dt}\right|_{m,\rho}$$
$$= (1-f_{red})\{\lambda_m + \kappa_+ + f_{red}(\alpha+\beta) \, sSFR\} \quad (7)$$

where $\lambda_m$ is the mass-quenching rate given by equation (17) in P10 as a constant μ times the instantaneous star-formation rate *SFR* and α is the logarithmic slope of the mass function at mass *m* and is given by equation (9) in P10, i.e., α=(1+$\alpha_s$)−m/M*(as distinct from the Schechter faint end slope $\alpha_s$). M* is the characteristic mass of the Schechter function. $\kappa_+$, as defined in P10, is the influx of new red galaxies into the mass bin per unit time due to major mergers and is assumed to be zero in the most under dense regions.

In P10, it was shown that, if β ~ 0, the red fraction in low density environments $\rho_0$ (where environmental quenching is negligible and $\kappa_+$ ~ 0) gradually approaches a "steady-state" in which $df_{red}/dt = 0$ at a given *m*. In this condition, the influx of blue star- forming galaxies which are brought up from lower masses by their star-formation, is matched by the production of new red passive galaxies via quenching, producing no change in the red fraction across all masses. The "steady-state" red fraction is therefore given by setting the derivative in equation (7) to zero, i.e.

$$f_{red}(m,\rho_0) = \varepsilon_m = \frac{1}{1-(1+\alpha_s+\beta)\frac{M_*}{m}} \quad (8)$$

We also refer the reader to Appendix B for an alternative derivation of equation (8). This is over-plotted in Fig. 3 as the dark orange line, using the usual Schechter parameters determined in P10 from SDSS sample of $M^* = 10^{10.6} M_\odot$ and $\alpha_s$ = -1.4 and setting $\beta = 0$.

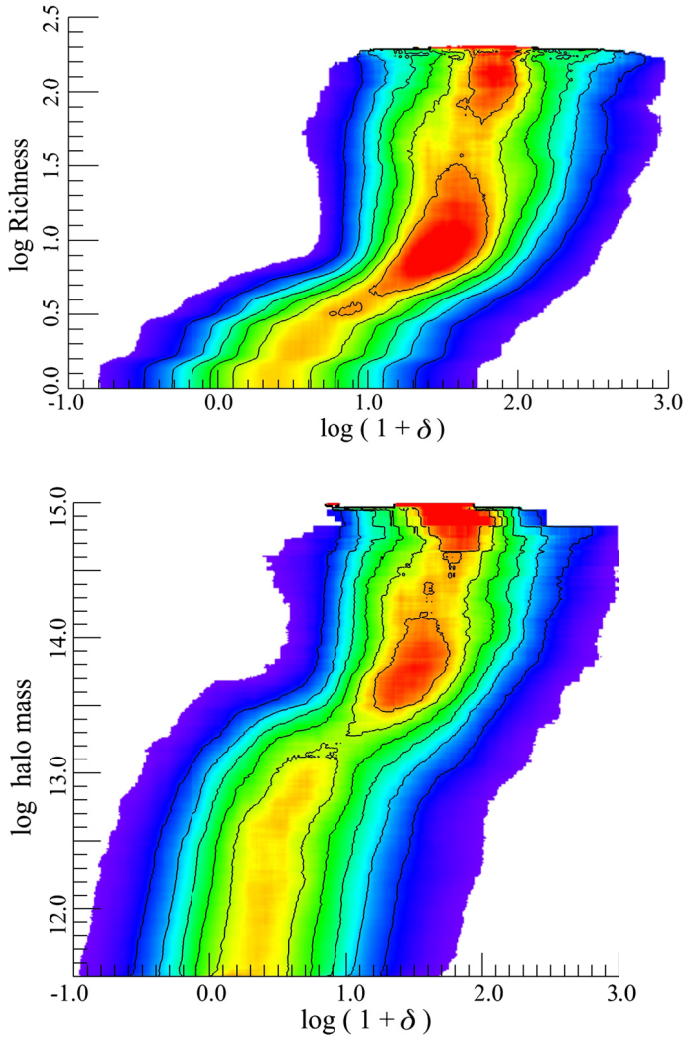

Figure 4. The distribution of 5th nearest neighbor over-densities for the galaxies (both centrals and satellites) in the group catalogue as a function of the richness (top panel) of their groups. The distributions are normalized to unity for a given richness. Both the richness and the over-densities are defined using the same set of tracer galaxies. As a result, the strong correlation for low richnesses log R < 0.7 (R < 5) reflects the fact that the 5th nearest neighbor will lie outside the group and the over-density will reflect the group richness rather then the position within the group. At higher richnesses, where there are many more than 5 members, the 5th neighbor is likely to be within the group, and the density estimator is therefore more sensitive to the radial location of the galaxy within the group, or cluster, rather than the overall richness, destroying this correlation. The distributions of galaxies in this figure should be born in mind in Fig. 6. We also show the distribution as a function of halo mass in the lower panel. In this case, the halo masses are defined using different set of tracer galaxies from the over-densities.

The fact that in Fig. 3 the red fraction determined from equation (8) (the dark orange line) well traces the observed red fraction of centrals in this paper (the blue line) and the red fraction in the low density D1 quartile from P10 (green line) strongly supports the simple picture for the evolution of galaxies that we developed in P10 and in the current work.

## 5. GROUP RICHNESS AND GALAXY ENVIRONMENT

The analysis in Section 4 has used the over-density δ as an environmental parameter, coupled with the identification of galaxies as centrals or satellites. This was to provide a direct link to the analysis in P10. This over-density is based on the distance to the 5th nearest neighbor that lies above a given luminosity threshold.

The dark matter halo mass is generally assumed to be a fundamental driver of galaxy evolution. We would expect a reasonably good proxy of halo mass on group scales to be given by the group richness, i.e. the number of members above some luminosity or stellar mass threshold (see e.g. Knobel et al. 2009). In this section we first look at the correlation between these two environment estimators, and then try to determine which of these is the main driver of the satellite quenching process discussed above.

### 5.1 The correlation between over-density and group richness

Fig. 4 shows the distribution of observed over-densities and parent group richness for all galaxies in the group catalogue. The distribution is determined first by applying a sliding box of width 0.3 dex in over-density and 0.3 dex in richness, and then re-normalized to unity at a given richness. There is a clear correlation between the richness and the over-density at log R <

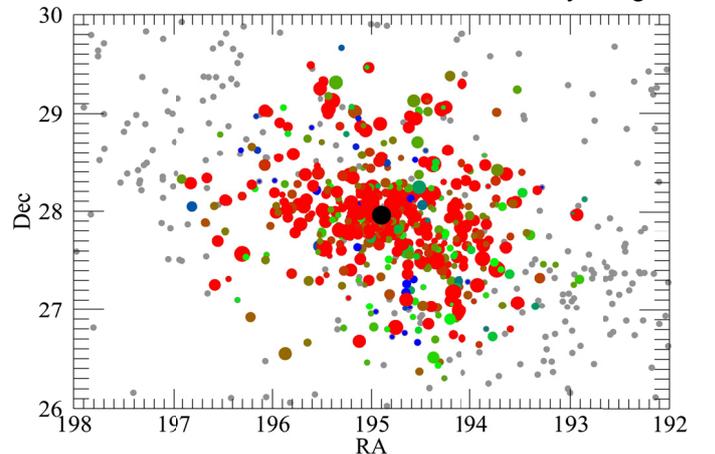

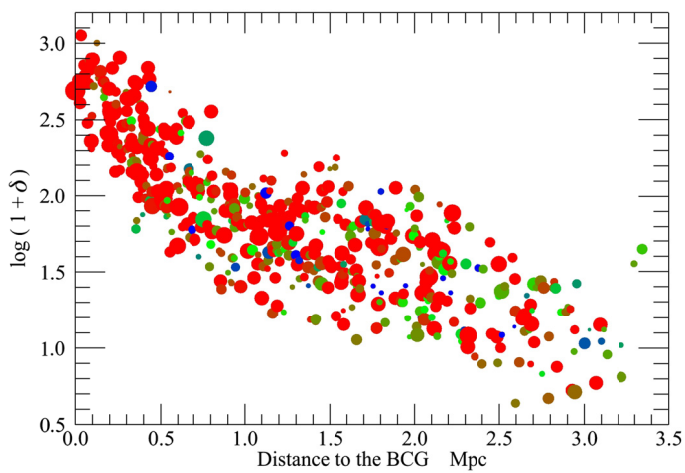

Figure 5. The spatial distribution of galaxies in the Coma Cluster in SDSS as an example of one rich group (top panel) and the over-density of each group member as a function of its distance to the BCG (Bright Central Galaxy) (lower panel). In both panels the size of each round dot is scaled according to the stellar mass of the galaxy and is also color coded with its $(U-B)_{rest}$ color. In the top panel, the black dot in the center is the BCG, which is also the most massive galaxy in the group. The grey dots are the background galaxies that are not cluster members.

0.7. Above this point, the correlation becomes much weaker. This change of behavior is almost certainly linked to the definition of over-density: for poor groups, the 5th nearest neighbor will lie beyond the group boundary, and the over-density will be closely related to the richness of the group and not to the location of a galaxy within the group. For richer groups, the 5th nearest neighbor distance is sensitive to the position within the group (or cluster - we use the terms interchangeably here). In other words, the transition between the two regimes apparently simply depends on the choice of $N$ in the $N$th nearest neighbor estimation of density.

The dependence of $\delta$ on cluster-centric radius for rich systems is illustrated in Fig. 5, which shows the projected spatial distribution of the Coma cluster members and the over-density $\delta$ of each member as a function of the distance to the Brightest Central Galaxy (BCG). As anticipated, there is a strong anti-correlation between the over-density and the distance to the BCG. We repeated this analysis with some other rich groups randomly drawn from the SDSS group catalogues and find similar results.

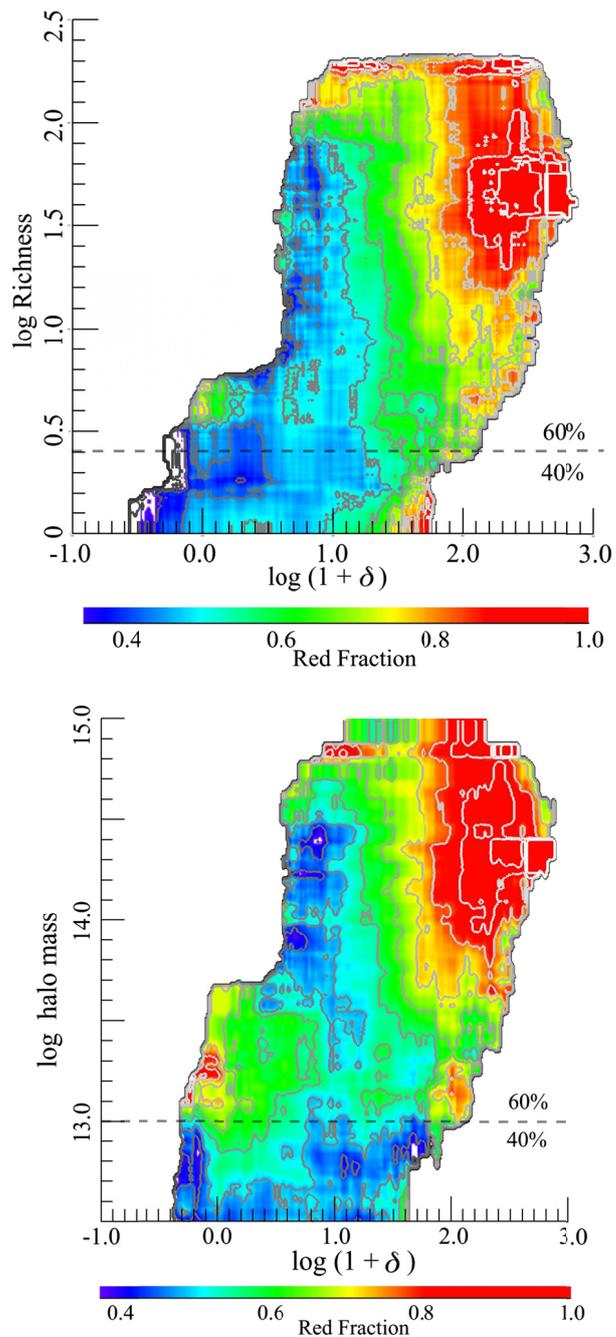

Figure 6. The red fraction of the satellite galaxies with stellar mass in the narrow mass range of $9.5 < \log m/M_\odot < 10.0$, as a function of over-density and group richness (top panel). The contour lines of the red fraction are essentially vertical, which implies that, at fixed mass and over-density, the red fraction doesn't depend on group richness, while at a given richness the red fraction steadily increases with increasing over-density. The distribution of over-densities for galaxies in groups of a given richness is given in Fig. 4. The dashed line marks the richness above which 60% of satellites are found. We also show the red fraction as a function of over-density and halo mass in the lower panel. Similar trends of the independence of red fraction on the halo mass are seen, especially in the more massive halos.

### 5.2 The relative effects of over-density and group richness on the satellite-quenching

To disentangle the influence of over-density and group richness on the satellite quenching process, we examine the red fraction of satellites at a fixed stellar mass as a function of the galaxies' over-densities and the parent group richnesses. In order to minimize variations of galaxy mass, and thus remove the variable effects of mass-quenching, whilst maintaining a reasonable number of galaxies for good statistical accuracy, we select satellites only within a limited range of stellar mass, $9.5 \leqslant \log m/M_\odot \leqslant 10$. This is the mass range where the effects of environment quenching are most apparent in P10.

The result of this exercise is shown in Fig. 6. Overall, the form of $f_{red}$ on Fig. 6 reflects the environment-quenching efficiency of P10. The $f_{red}$ increases monotonically with $\delta$, but especially steeply at $\log \delta > 1.0$ (see Fig. 8 in P10). It should be recalled that these $\delta$ are locally projected over-densities so the associated physical over-densities will be substantially higher.

As noted above, the richness $R$ and over-density $\delta$ are correlated for the lowest richness groups, where the majority of satellite galaxies reside, but decouple for the richer ones where $\delta$ indicates the radial location of the galaxy within the group halo. The striking vertical contours of $f_{red}$ in the $\delta$-richness plane (Fig. 6) shows clearly that, at least for these richer systems, the over-density is the primary determinant of the satellite quenching process, rather than the richness that is presumably a proxy for the overall halo mass. Although many galaxies inhabit the lowest richness systems, where our definition of Richness is most severely quantized in log space, and where $\delta$ and Richness are more strongly correlated, it should be noted that 60% of satellites lie above the dashed line in Fig. 6, i.e. with $R \geq 3$, where the independence of red fraction with Richness is most clearly demonstrated.

We note that van den Bosch (2008b) has previously pointed out the independence of average color for the satellites on the halo mass at fixed stellar mass, as seen here, but also claimed that this was independent of the location within the halo. We suspect that this latter statement is inconsistent with the results presented here, and also by extension those in P10.

The fact that the local over-density, broadly interpretable as the location within the halo, is more important than the overall halo mass in controlling the evolution of satellites gives important clues as to the physical nature of the satellite quenching process. Obvious possibilities involving the local density would include strangulation (Larson, Tinsley & Caldwell 1980; Balogh, Navarro & Morris 2000; Balogh & Morris 2000, Feldmann et al 2010) in which the galaxy is deprived of its immediate gas supply in its halo, ram-pressure stripping (e.g., Gunn & Gott 1972; Abadi, Moore, & Bower 1999; Quilis, Moore & Bower 2000) in which the interstellar medium is removed, and tidal stripping and harassment (Farouki & Shapiro 1981; Moore et al. 1996). Local over-density can also reflect the cosmic epoch at which a satellite became a satellite, i.e. the length of time that a galaxy has been a satellite.

In assessing these possibilities, an important constraint comes from the fact that the action of satellite quenching manifestly does *not* depend on the stellar mass of the galaxy involved (see Figs. 2 and 3). Furthermore, the star-formation rates in those satellites that have *not* been quenched is found to be very similar to the star-formation rates in the star-forming centrals of the same stellar mass. This is illustrated in the two panels of Fig. 7, in which the sSFR of the star-forming centrals and satellites are plotted. These are within 0.15 dex over the full stellar mass range and are essentially identical in the mass range of most interest at $\log m/M_\odot > 10.0$.

### 5.3 The quenching of satellite galaxies in different mass haloes

In the previous section we interpreted our observed Richness as a proxy for the dark matter masses of the haloes in an effort to identify the environmental drivers of our "environment-quenching" process. We concluded that environment-quenching appeared to be driven by the over-density $\delta$ rather than by the richness of the group or the parent halo-mass. One advantage of using the observed Richness is that it can be defined using exactly the same tracer galaxies as our over-density estimator.

In this Section we explore further whether there are linkages between the quenching of galaxies and the masses of their parent haloes. We use for this purpose the halo masses as defined by Y07. As described in Section 3.3, these halo mass estimates are based on the integrated stellar masses of the groups, calibrated against mock catalogues.

Fig. 8 shows the stellar masses of individual galaxies in our sample plotted against the Y07 dark matter mass of their parent halo. Star-forming central galaxies (upper left panel) show a tight correlation with the parent halo mass. Although this undoubtedly reflects to a certain degree the method of computation of halo mass via the integrated stellar mass, such a correlation is also expected if the star-forming centrals have been built up through the conversion of some fraction of the baryonic mass in the haloes into stars.

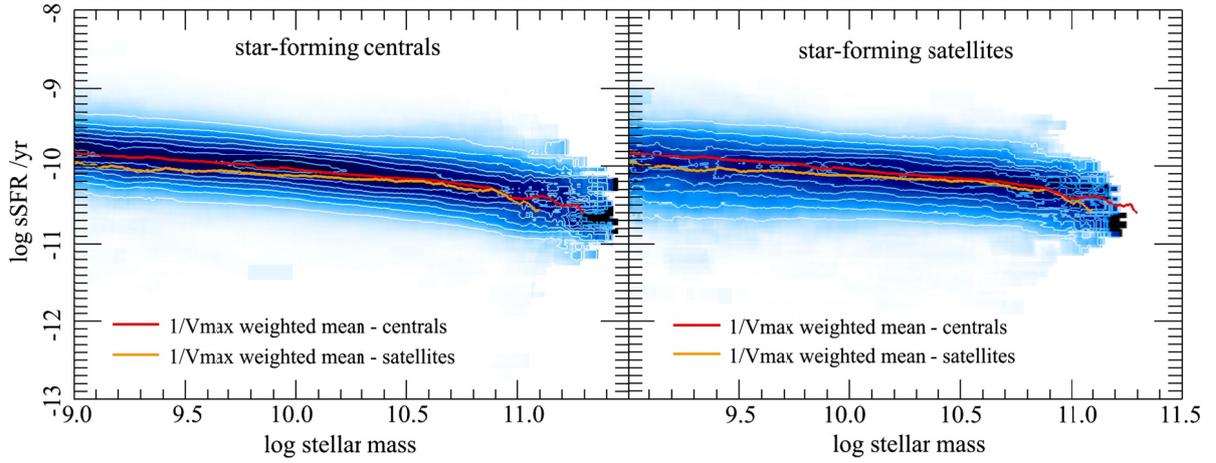

Figure 7: The relationship between sSFR and stellar mass for SDSS DR7 star-forming central galaxies (left) and star-forming satellite galaxies (right). The red and dark orange lines in each panel show the $1/V_{max}$ weighted mean of the sSFR for the star-forming centrals and satellites respectively. These two curves very similar, especially at the more massive end, $\log m/M_\odot > 10.0$, indicating that the SFR of star-forming galaxies doesn't depend on the central and satellite dichotomy.

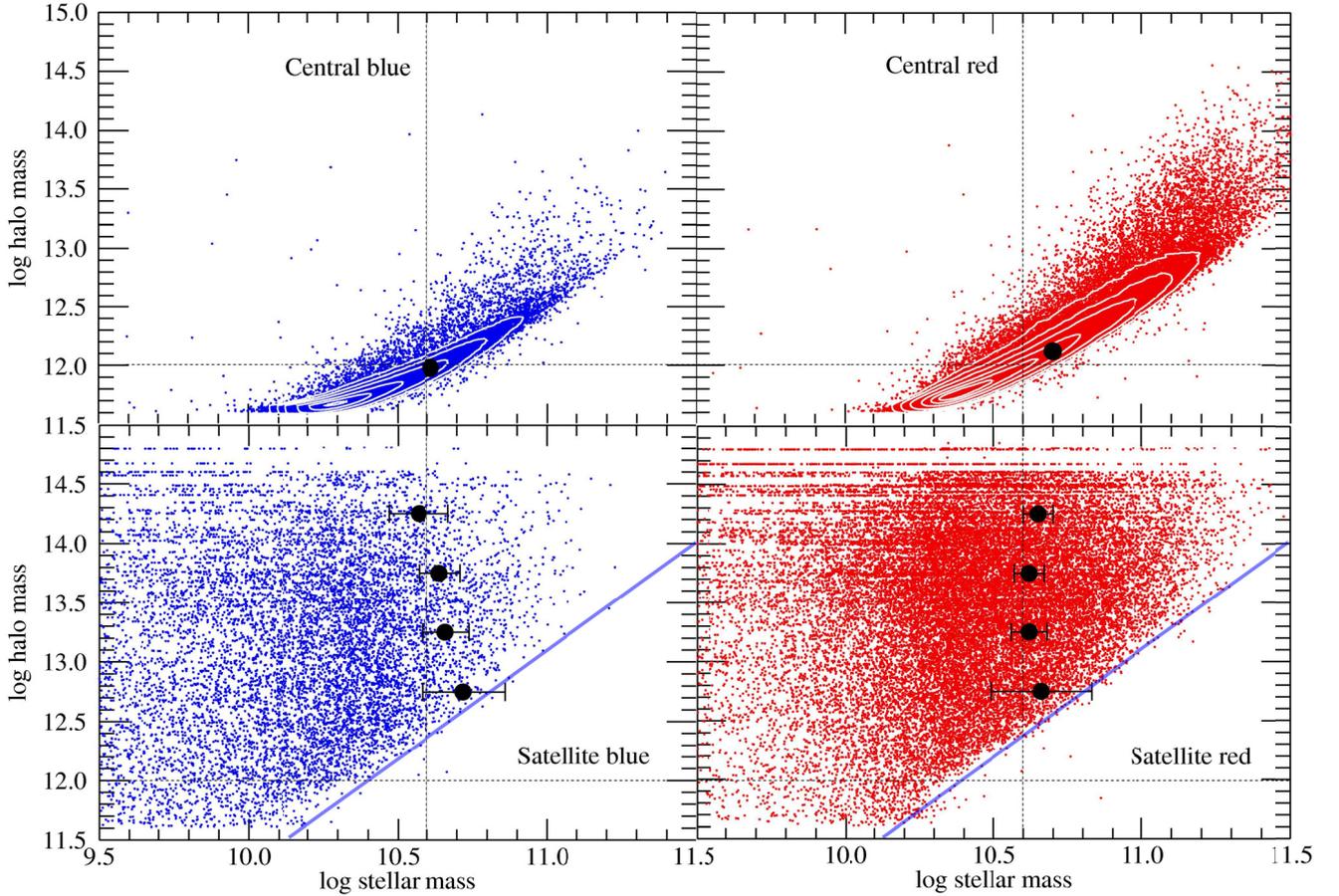

Figure 8: Stellar masses and parent halo masses for red and blue central and satellite galaxies. Blue central galaxies show a tight correlation between stellar and halo mass and a fall-off in number above a stellar mass of $10^{10.6}$ $M_\odot$, corresponding to M* in the mass-function, which may be associated with a halo mass of about $10^{12}$ $M_\odot$ (the two mass scales indicated by the dotted lines). For these galaxies it is difficult to distinguish between stellar mass and halo mass. The diagonal light blue lines in the two bottom panels show the maximum satellite stellar mass that is a consequence of the tight correlation between halo mass and the central stellar mass (since satellites must be less massive than centrals), estimated from the sharp boundary of the galaxy distribution. The black dots in the two top panels show the M* of the Schechter function fitted for all blue and all red central galaxies respectively (from table 2). The black dots in the two bottom panels show the M* of the Schechter function fitted at given halo mass for blue and red satellite galaxies respectively (from Figure 9 & 10). Blue satellite galaxies show identical M* over more than two decades of halo mass, indicating that the action of mass-quenching is independent of (today's) halo mass. Red satellite galaxies also show similar M* as the blue satellite galaxies over two decades of halo mass, as predicted from our model (also see section 6.1) and also supporting that mass-quenching is independent of halo mass. The distribution of masses of red centrals (upper right panel) reflects the survival of centrals as their halo masses increase (see text).

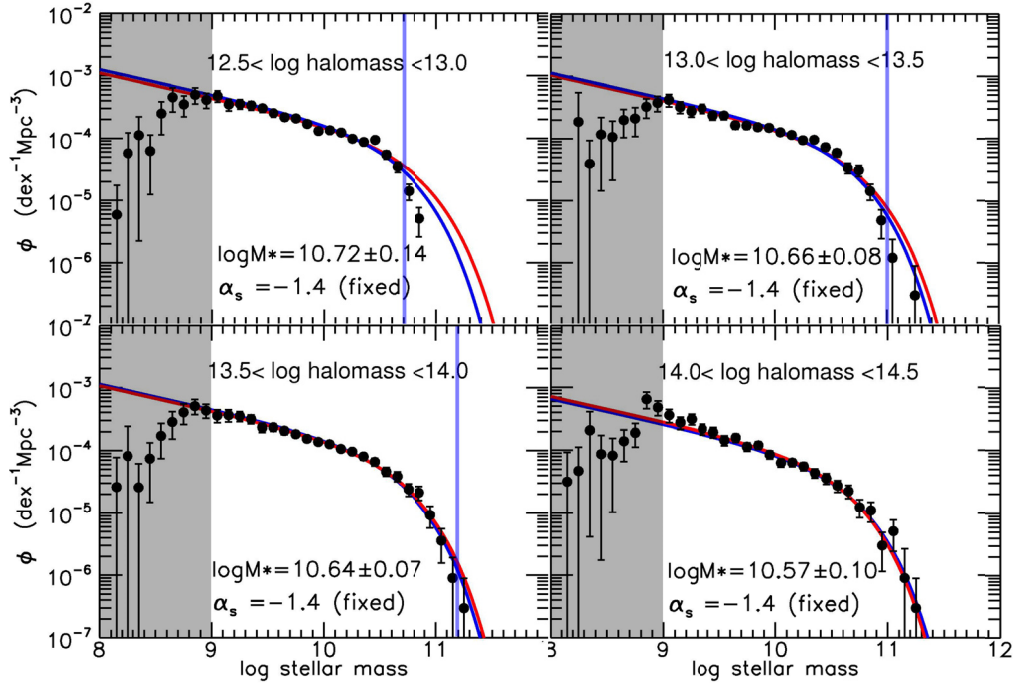

Figure 9. Galaxy Stellar Mass Function (GSMF) of SDSS star-forming satellite galaxies at different parent halo mass. In each panel the observed data is shown as black dots. The grey shaded regions on the left show where the data is seriously incomplete. The light blue vertical line in each panel shows the maximum satellite stellar mass (at given halo mass), the value of which is estimated from the diagonal blue lines in the lower panels of Figure 8. In order to compare M*, a single Schechter function is fitted to the data above the complete mass of ~$10^9$ M$_\odot$ and below the maximum satellite stellar mass (i.e. to the left of the vertical blue line), assuming fixed $\alpha_s$=-1.4. The best fit Schechter function is plotted as the red line and the M* is labeled in each panel and it provides a fully satisfactory fit to the data within the valid stellar mass range. The mass function of star-forming galaxies in the most under dense D1 density quartiles determined in P10 is renormalized (in number density $\phi$) and is plotted as the blue curve in each panel. It's evident that the shape of the Schechter function, in terms of $\alpha_s$ and M*, is identical across a broad range of halo mass and is also very similar to the mass function of star-forming galaxies in the most under dense D1 density quartiles.

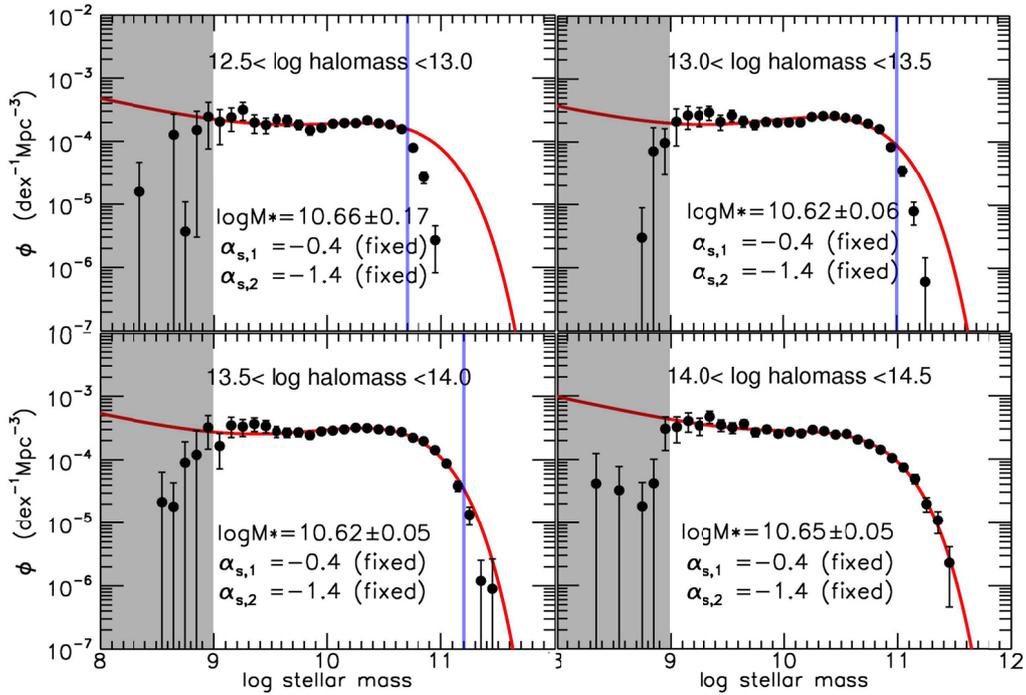

Figure 10. Similar to Figure 9, but for passive satellite galaxies. In order to compare M*, following the P10 formalism, a double Schechter function is fitted to the data above the complete mass of ~$10^9$ M$_\odot$ and below the maximum satellite stellar mass (i.e. to the left of the vertical blue line), assuming fixed $\alpha_{s,1}$= -0.4 and $\alpha_{s,2}$= -1.4 in all panels. This provides a fully satisfactory fit to the data within the valid stellar mass range. The M*, is largely independent of the halo mass and is identical to the M* of the star-forming satellites and thus also to the M* of the star-forming galaxies in the most under dense D1 density quartiles.

The redshift evolution of the *sSFR* of star-forming galaxies evolves roughly as $t^{-2.2}$ (Pannella et al. 2009, P10, Elbaz et al. 2011) with a shallow logarithmic slope of the *sSFR–m* relation with a value of $\beta$ around -0.1 (e.g. Daddi et al. 2007; Elbaz et al. 2007; Pannella et al. 2009). Although steeper values, with $\beta \sim -0.4$ have also been claimed (Noeske et al 2007, Karim et al 2011, Rodighiero et al 2010), the last of these has now been revised to a much shallower $\beta \sim -0.2$ (Rodighiero et al 2011). The discrepancy of the observed values of β is discussed, for example, in Stringer et al. 2011.

The mean specific dark matter halo accretion rate in dark matter simulations also increases in a similar way, approximately as $(1 + z)^{2.2}$ (Genel et al. 2008, McBride et al. 2009) and it also depends only weekly on the halo mass, with an equivalent $\beta \sim 0.1$. The similarity of these two behaviors suggests that the build-up of stellar mass is likely closely related to the mass increase of the parent haloes. The slope of this $m_{halo}$-$m*$ relation is evidently curved, being steeper than unity at high masses and shallower at low masses, reflecting the difference in faint end slopes $\alpha_s$ of the halo and galactic stellar mass functions.

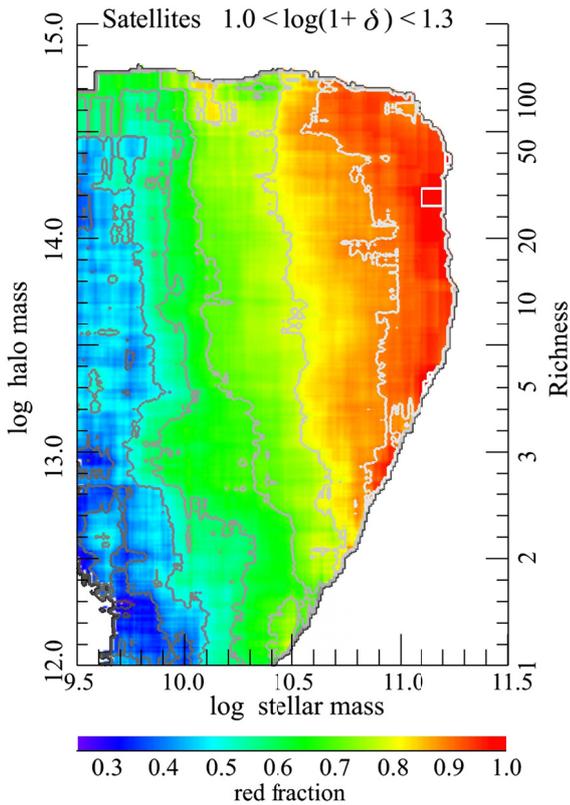

Figure 11: The red fraction of satellite galaxies within a limited range of over-density as a function of stellar mass and parent halo mass. By limiting the range of over-density we eliminate the effects of environment-quenching and isolate those of mass-quenching. The effect of mass-quenching is evidently independent of (today's) dark matter halo mass over more than two decades of halo mass.

The fall-off in the number of blue centrals at log $m/M_\odot >$ 10.6 is due to the exponential cut-off in the mass-function above M*. In P10, we showed that this cut-off could be naturally understood as a result of our mass-quenching law, and pointed out that setting $\eta = \mu$ SFR for the mass-quenching rate is entirely equivalent to imposing a survival probability for a galaxy to attain a mass $m$ of $P(m) = \exp(-\mu m)$, i.e. implying a *de facto* limit to the stellar mass of the galaxies. Given the tight correlation between stellar mass and halo mass for star-forming central galaxies, any "stellar-mass-limit" of this form could then be associated with a limit to the mass of dark matter haloes that are able to support star-formation, as has been argued on theoretical grounds, e.g. through shock heating of the gas in massive halos ($>10^{12}$ $M_\odot$) (Kereš et al. 2005, Dekel & Birnboim 2006) or gravitational gas heating ($>10^{13}$ $M_\odot$) (Khochfar & Ostriker 2008; Dekel & Birnboim 2008). This connection between the quenching-rate-approach of P10 and mass-limit-approach is illustrated by the dashed lines in the upper left panel of Fig. 8. The halo mass corresponding to the observed $M^*$ of the star-forming galaxies is of order $10^{12}$ $M_\odot$ corresponding closely to the limit favored by Dekel and Birnboim (2006). The close link between the stellar mass and halo mass for the star-forming *centrals* also makes it difficult to determine, for these galaxies, which actually controls the mass-quenching.

The star-forming satellite galaxies occupy, as would be expected, a much broader range of halo masses (see figure 8, lower left panel) than the centrals, extending up from the tight relation between stellar and halo mass exhibited by the latter (upper left panel). It is however immediately clear from the lower left panel that these star-forming satellite galaxies exhibit a very similar Schechter M* (plotted as the black dots), over at least two decades of halo mass. The mass functions and the Schechter fits of the star-forming satellite galaxies in different ranges of halo mass are shown in Figure 9. In each panel the observed data is shown as black dots. The grey shaded regions on the left show where the data is seriously incomplete (i.e. below $\sim10^9$ $M_\odot$). The light blue vertical line shows the maximum stellar mass limit that is allowed for the star-forming *satellite* at given halo mass given the tight correlation between halo mass and the stellar mass of the *central* galaxy, given that the satellite is required to be less massive than the central. This mass limit is estimated from the sharp boundary of the galaxy distribution in the lower two panels of Figure 8, and is shown in these panels as the diagonal blue lines. A single Schechter function is fitted to the data in the range of stellar mass above the mass completeness limit and below this maximum mass limit. In order to make direct comparison of the M*, the Schechter function is fitted with fixed $\alpha_s$=−1.4 in all halo mass bins and this has provided a fully satisfactory fit to the data within the valid stellar mass range. The same exercise has been applied to the passive satellites and the results are shown in Figure 10, except that a double Schechter function is now used for the fit with fixed $\alpha_{s,1}$= −0.4 and $\alpha_{s,2}$= −1.4 (see P10 and Section 6.2 in the current paper for details). Again, this has provided excellent fits to the data within the valid stellar mass range.

A key result is that it can clearly be seen that the shape of the Schechter function of the star-forming satellites is identical across a broad range of halo mass, especially, in the free parameter M*, and also in $\alpha_s$. This shape is also essentially identical to the shape of the mass function of *all* star-forming galaxies in the most under dense D1 density quartiles shown by the blue curves in Figure 9 (with arbitrary renormalization in number density). The shape is also identical to the one of the star-forming centrals (Section 6.2) with, again, essentially identical M*.

In our formalism, the shape, and especially the value of M*, of the mass function of surviving star-forming galaxies is a direct consequence of the action of the mass-quenching process. Indeed, in many regards, we can regard "mass-quenching" as that process, whatever its physical origin, that establishes the value of M*. The fact that M* is independent of halo mass for satellites clearly shows that mass-quenching acts independently of the parent halo mass for these galaxies, at least in the range above $10^{12}$ M$_\odot$. Satellites in the most massive $10^{15}$ M$_\odot$ haloes are evidently no more likely to be *mass-quenched* than those in $10^{12}$ M$_\odot$ haloes.

It is noteworthy how the underlying "universal" Schechter functions for star-forming and passive galaxies that is produced by the mass-quenching process emerges from underneath the maximum mass for satellites (the vertical line in Figures 9 & 10) as that limit moves to higher stellar masses as the halo mass increases.

In addition to the mass function, we can also look at the red fraction of the satellites in a limited range of over-density $\delta$. As shown in Fig.4, there is a correlation between halo mass and mean $\delta$. In P10 we showed that the environment quenching acts on $\delta$. Therefore by limiting the range of $\delta$ we eliminate the effects of environment- quenching and isolate the effects of mass-quenching. This is analogous to the analysis of the environment-quenching of satellites in Figure 6, where we eliminate the effects of mass-quenching by looking at a narrow range of mass and isolate the effects of environment-quenching. We choose an interval of $1.0 < \log(1+\delta) < 1.3$ which contains a large number of satellites (see Fig 1) and spans the full range of Richnesses and halo masses (see Fig. 6). This is shown as a function of stellar and halo mass in Fig. 11. The red fraction is strikingly independent of halo mass while depending strongly on stellar mass - echoing an earlier conclusion of van den Bosch (2008a - but see above discussion about the dependence on over-density).

We can therefore see that three of the most important aspects of satellite galaxies existence, i.e. their star-formation rates, and the action of both mass- and environment- quenching, do not appear to depend on the mass of the parent halo. The relevant observations are

(i) the independence of the sSFR of star-forming satellites on parent halo mass (in Fig. 12);
(ii) the independence of M* for satellites on parent halo mass (Fig 8, bottom left hand panel, Fig. 9 & 10) since this reflects the action of mass-quenching;
(iii) the independence of satellite-quenching on halo mass shown by the independence of the red-fraction of satellites on richness and halo mass at fixed stellar mass and over-density (as shown in Fig. 6).

### 5.4 The quenching of central galaxies

For central star-forming galaxies, due to the tight correlation between their stellar mass and their parent halo mass (Fig 8, top left panel), it is very difficult to differentiate whether the stellar mass or the halo mass dominates the (mass-)quenching of the galaxies.

Nevertheless, as we stressed above, the shape of the mass function of the (surviving) star-forming centrals is a key diagnostic of the action of mass-quenching. The fact that the star-forming centrals have exactly the same value of M* as the population of satellites suggests that satellites and centrals suffer the same physical process in mass-quenching. The fact that M* for satellites is independent of halo mass above $10^{12}$M$_\odot$, and thus that the mass-quenching process for satellites is independent of halo mass, then suggests that this is also likely to be the case for the centrals.

This is an indirect argument, but it would be strange if the mass-quenching process that produces M* were to operate independently of halo mass for satellites, but was driven by halo mass for centrals, and yet the two processes produced exactly the same value of M*.

How could one save the popular idea that halo mass is in fact the driver of mass-quenching of satellites, and therefore by extension, of centrals also? One way would be to simply assert that the total parent halo mass was irrelevant and that all satellites retained complete "sub-haloes" which control the gas flow onto the satellites, independent of the larger parent haloes in which they reside. We suspect that this is implausible physically (see e.g. Hayashi et al. 2003, Gao et al. 2004, Kazantzidis et al. 2004). Alternatively, we could imagine that the increase in halo mass, which evidently moves star-forming satellites off the tight halo-stellar mass correlation of the centrals, takes place so recently that there has been negligible increase in the stellar mass of the satellites through ongoing star-formation. If the growth in stellar mass has been small while the galaxies are satellites, then the masses and red-fractions of the satellites in these high mass haloes will actually reflect the operation of mass-quenching before their infall, i.e., while the satellites were still centrals in their own (lower-mass) haloes. However, if satellite galaxies are increasing their masses significantly after their infall, then the only way that they will end up with the same M* as shown in the lower panels in Fig. 8 is if mass-quenching acts in the same way *inside* the larger halo as *outside*. Indeed as shown in Fig. 7, the satellites have essentially the same sSFR as central galaxies, while Fig. 12 shows that the mean <log SFR> is independent of halo mass. Fig. 1 & 2 in P10 also show that the mean <log SFR> is independent of environment (also see von der Linden et al. 2010), all suggesting that some significant mass increase (and thus attendant mass-quenching) will have taken place since the galaxies became satellites.

We regard the above arguments as providing evidence against the popular idea that dark matter halo mass controls the evolution of galaxies. Not least, the invariance of mass-quenching with parent halo mass suggests to us that more internal processes, which somehow know about the stellar mass (or star-formation rate), are involved in limiting the mass growth of galaxies.

In closing we point out the difficulty of interpreting red fraction as a diagnostic of quenching in different halo masses. For instance, there is evidence that at fixed stellar mass, the red fraction of centrals increases with increasing halo mass. As discussed at the end of Section 5.5, this need not imply that the halo mass controls the quenching of centrals, since it would also arise in a very natural way in a scenario in which stellar mass controlled quenching.

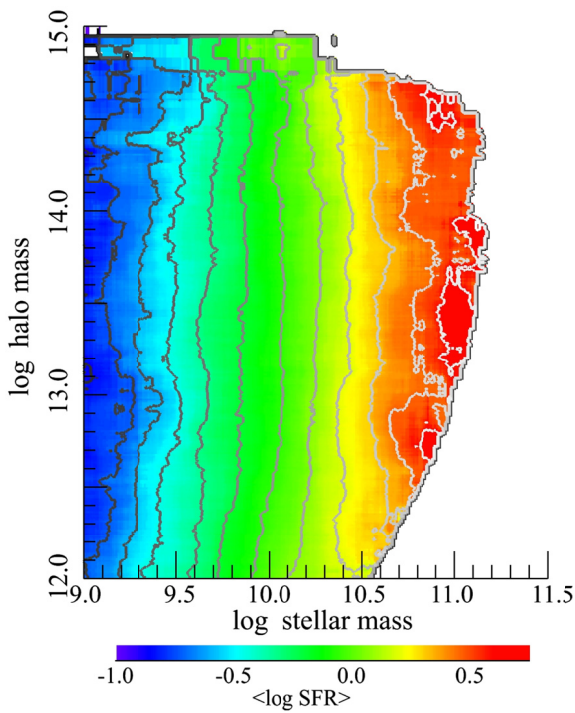

Figure 12: $1/V_{max}$ weighted mean <log SFR> as a function of stellar mass and halo mass for star forming satellite galaxies in SDSS DR7. The contour lines of the <log SFR> are essentially vertical, which implies that the independence of SFR on halo mass at given stellar mass for star forming satellites.

## 5.5 The masses of central galaxies as a function of halo mass

The interpretation of the right hand panels in Fig. 8 is more subtle. Once star-formation in a galaxy is quenched, the stellar mass of the galaxy should not increase further, except through the effects of post-quenching merging.

We would therefore expect central galaxies to grow along the star-forming "main-sequence" and to be quenched at some point following the exponential survival probability $P(m)$. Galaxies might then be expected to move vertically in the upper left panel of Fig. 8 as their dark matter haloes continued to increase in mass without attendant star-formation.

This increase in halo mass will occur as haloes merge with other haloes, large or small. As a result, not all of the quenched centrals will remain centrals. Each time two haloes merge, only the more massive of the two centrals will remain as the central of the new halo. The other will become a satellite, or may even disappear completely if it merges with the larger galaxy. One way or the other, one of the two centrals will be removed from this panel of the diagram. Clearly the chance of a central to survive as a central will depend on its stellar mass, and thus on the mass of its halo when it was quenched. As a result, if we start with a distribution of masses of quenched centrals, then statistically, only the most massive ones will survive as centrals following substantial dark matter growth in the masses of their parent haloes. The distribution of the "surviving" passive centrals that lie above the main locus in Fig 8 will therefore be expected to be a triangular region, since low mass passive centrals would not be expected to "survive" a large increase in the dark matter mass. Indeed, this could explain that the limited vertical scatter of the $M_h$ - $M^*$ sequence of the central star-forming galaxies over much of the lower mass range. It is evidently about 0.3 dex. If a halo increases by more than a factor of two, then the central galaxy may well no longer qualify as the central of the resulting halo.

It is clear by comparing the upper two panels of Figure 8 that observationally the red fraction of centrals will increase with halo mass, at a given fixed stellar mass, simply because of the broader spread in halo mass for red galaxies than for blue galaxies. This in turn is a simple consequence of the halo growth that continues to occur after quenching of the star-formation and resulting constancy of the stellar mass.

This shows that a scenario in which quenching is a result of stellar mass alone will inevitably produce the possibly counter-intuitive result of a correlation of red fraction with halo mass at fixed stellar mass. Likewise, red centrals would be expected to have more satellites than blue ones.

As in P10, we emphasize the power of the shape of the mass-function of surviving star-forming galaxies as a diagnostic of quenching over the more usual red fraction approach.

## 6. THE MASS FUNCTIONS OF CENTRAL AND SATELLITE GALAXIES

In Sections 4 and 5 we developed the argument that the (stellar mass-independent) environment-quenching of P10 is actually a satellite quenching process that depends strongly on local over-density, but not on halo mass or on stellar mass. The average quenching efficiency is $<\varepsilon_{sat}> \sim 40\%$ when averaged over all environments.

In this section, we follow the P10 approach to derive the analytical forms and interrelationships of the galactic stellar mass functions (GSMF) that we would expect for star forming and passive galaxies that are either centrals or satellites. These will then be tested against observed mass functions that we construct from the SDSS group catalogue.

### 6.1 Predictions for the mass-functions

We first recap and refresh the simple arguments given in P10 for the mass-functions of the passive galaxies that will be produced by our different quenching processes. The mass-function of newly-formed passive galaxies will be simply obtained by multiplying the mass-function of star-forming galaxies by the appropriate quenching rate.

For mass-quenching, the quenching rate is proportional (see P10) to the SFR, i.e. to $m^{(1+\beta)}$. If the star-forming mass function is a Schechter function with some characteristic $M^*$ and faint-end slope $\alpha_s$, as expected from the P10 model, then the new passive galaxies will have also a Schechter function with exactly the same $M^*$ but with a modified faint-end slope with $\Delta\alpha_s = 1$. For any mass-independent quenching process, such as the environment-quenching of P10, or the equivalent satellite quenching of the current paper, the mass-function of the new passive galaxies will again be a Schechter function, but now with the same $\alpha_s$ as well as the same $M^*$.

If the $M^*$ and $\alpha_s$ of the star-forming population stay the same over long periods of time, as they appear observationally to do

(see P10 and references therein), then the above statements about M* and $\alpha_s$ for the "new" passives will of course hold for the entire population of passive galaxies, leading to the same predictions for the final population seen today. This avoids the need to do a complicated integral over time (see the Appendix B for a simple demonstration).

With reference to P10, the mass function of the *centrals* should therefore be exactly the same (apart from the normalization $\phi^*$) to the mass function of all galaxies in the *lowest* density environments D1, but possibly with some additional effects of post-quenching merging (as discussed in P10). Neglecting for the moment the effect of merging, we should thus have a single Schechter function for the star-forming galaxies and a single Schechter function for the passive galaxies, both with the same *M*\* but with $\Delta\alpha_s = 1$. From the definition of $\varepsilon_m$ (in terms of $f_{red} = \phi_{red} / (\phi_{red} + \phi_{blue})$), the mass-function of the mass-quenched passive centrals can be determined from the mass-function of the star-forming centrals:

$$\phi_{cen,red}(m,\rho,t) = \phi_{cen,blue}(m,\rho,t) \frac{\varepsilon_m(m)}{[1-\varepsilon_m(m)]} \quad (9)$$

As discussed in section 4.3, $\varepsilon_m$ is independent of environment and is effectively constant over cosmic time in a "steady-state" but strongly depends on mass as shown in equation (8). Alternatively, the mass-function of the mass-quenched passive centrals can be derived directly from the continuity equation of the star-forming centrals (as demonstrated in the Appendix B) and it produces the same result of equation B4 as equation (9).

The mass function of the *satellite* galaxies should look like the mass function computed in P10 for the high density environments D4, but possibly *without* the additional effects of merging. We again expect a single Schechter function for the star-forming satellite galaxies, but a double Schechter function for the passive satellites, with the primary component due to those satellites which were mass-quenched and the secondary component, which will dominate at lower masses, coming from satellite-quenched satellites.

It is easy to see from the definitions of $\varepsilon_m$ and $\varepsilon_{sat}$ that the two components of the mass-function of the passive satellites will be given, in terms of the observed mass function of actively star-forming satellites, $\phi_{sat,blue}$, and our quenching efficiencies, by:

$$\phi_{sat,red,1}(m,\rho,t) = \phi_{sat,blue}(m,\rho,t) \frac{\varepsilon_m(m)}{[1-\varepsilon_m(m)]} \frac{1}{[1-\varepsilon_{sat}(\rho)]}$$

$$\phi_{sat,red,2}(m,\rho,t) = \phi_{sat,blue}(m,\rho,t) \frac{\varepsilon_{sat}(\rho)}{[1-\varepsilon_{sat}(\rho)]}$$

$$= \phi_{sat,red,1}(m,\rho,t) \frac{[1-\varepsilon_m(m)]}{\varepsilon_m(m)} \varepsilon_{sat}(\rho)$$

(10)

Following the same convention of notation in P10, $\phi_{sat,red,1}$ is the mass quenched satellites and $\phi_{sat,red,2}$ is the environment quenched satellites. As discussed in section 4.2, $\varepsilon_{sat}$ is independent of mass but strongly depends on environment. $\varepsilon_{sat}$ is also expected to be independent of epoch. However, as addressed at the end of section 4.2, this does not imply that the effects of satellite-quenching on the galaxy population are unchanging with cosmic time, since the median over-density of satellite galaxy population is expected to increase with cosmic time.

Since *M*\* will be the same for all components, and because $\varepsilon_m(m)$ evaluated at $m = M^*$ should be just given by $-\alpha_s^{-1}$ (from equation (8), setting $m = M^*$ and $\beta = 0$) it follows that the values of $\phi^*$ of the different components of the passive population should be related as follows:

$$\phi^*_{cen,red}(\rho,t) = \phi^*_{cen,blue}(\rho,t) \frac{1}{[-\alpha_{s,cen,blue}-1]}$$

$$\phi^*_{sat,red,1}(\rho,t) = \phi^*_{sat,blue}(\rho,t) \frac{1}{[-\alpha_{s,sat,blue}-1]} \frac{1}{[1-\varepsilon_{sat}(\rho)]}$$

$$\phi^*_{sat,red,2}(\rho,t) = \phi^*_{sat,blue}(\rho,t) \frac{\varepsilon_{sat}(\rho)}{[1-\varepsilon_{sat}(\rho)]}$$

$$= \phi^*_{sat,red,1}(\rho,t) \varepsilon_{sat}(\rho)[-\alpha_{s,sat,blue}-1]$$

(11)

The ratio of the $\phi^*$ of the two components of the passive satellite population (in all environments) is given by the last equation, and for $\alpha_s \sim -1.5$ and the environment-averaged satellite quenching efficiency (independent of stellar mass) in SDSS: $<\varepsilon_{sat}> \sim 0.4$, and it should be about 5.0 in favor of the primary (mass-quenched) component.

In P10, we pointed out that any merging of the passive galaxies, *after* they have been quenched, will lead to an increase in stellar mass of these galaxies and to a loss of the equality of *M*\* between them and the star-forming galaxies that was established by the mass-quenching process. Recall that, if *M*\* for the star-forming galaxies is constant with time (as observed, see P10 for a discussion), then the passive galaxies should always have been initially produced with the same *M*\* as the star-forming galaxies, albeit with a different $\alpha_s$. Subsequent mass growth of the passives through merging will lead to a change in their *M*\*, breaking this equality with M* of the star-forming galaxies.

We explored in P10 a simple model involving nearly equal mass mergers of quenched galaxies in the primary (mass-quenched) passive population with $\alpha_s \sim -0.4$. This leads to an expectation of a correlated shift in *M*\* and $\alpha_s$ if the resulting composite mass function is fit by a single Schechter function. The shift in $\alpha_s$ arises because the galaxies are merging with others from the same population, so that a merger produces both a change in mass and a change in number density. In P10 we derived $\Delta\alpha_s = 1.6\,\Delta\log M^*$ from a numerical simulation and showed that there was observational evidence for this correlated shift within the passive population in the high density D4 quartile.

In the central-satellite scenario, we might expect most mergers to involve satellites merging with centrals. Satellite-satellite mergers would be quite rare (e.g. Angulo et al. 2009), and central-central mergers are impossible according to our definition, since by the time two centrals can hope to merge, one of them by definition loses its central status. Assuming that the merger probabilities are independent of stellar mass (see P10), the effect of merging on the centrals will therefore be to shift in mass alone, *without* an associated change in their number density (since the central stays being a central). We would therefore expect a shift in log M* with no change in $\alpha_s$. Under the same assumption, the effect on the satellites is simply to decrease their number density, again with no change expected in $\alpha_s$.

### 6.2 Observational tests

We test the above predictions directly by constructing the mass function of centrals and satellites using the SDSS DR7 group catalogue. These are shown in Table 2 and Fig. 13. In each panel of Fig. 13 we show the mass functions for star forming and passive galaxies for both centrals and satellites as determined from the data (in bins of 0.1 dex in mass). We also show the best fit Schechter functions above the mass completeness, which is around $10^9$ M$_\odot$ with $V_{max}$ weighting. Finally we show the predictions based on the previous section, normalized to the observed star-forming mass-function for each set of galaxies, i.e., we take the observed star-forming mass functions for centrals and satellites as inputs of equation (9) and (10) to predict the passive mass functions and the global mass functions (the sum of the star-forming and passive galaxies).

From equation (10), the mass function of the passive satellites is given by both $\varepsilon_m$ and $\varepsilon_{sat}$ and is thus, through the latter, a function of $\rho$. The former is calculated analytically using the continuity equation as in equation (8) and also in equation (B6), and uses as inputs $M^*$ and $\alpha_s$ from the *blue* population only. At present, $\varepsilon_{sat}$ is determined observationally from the measured red fractions of satellites (shown in Figure 2) with the additional constraint that it is independent of stellar mass. $\varepsilon_{sat}$ will be related to the density field construction method, structure growth and satellite quenching rate. For a given satellite population with some distribution of $\rho$, we could calculate the mass function as a function of $\rho$ and then add all these mass functions together according to the $\rho$ distribution. Alternatively, we could also use an average $<\varepsilon_{sat}>$ that was obtained by weighting $\varepsilon_{sat}(\rho)$ by the $N(\rho)$ distribution of the satellites. In fact, as a convenience, we approximate $<\varepsilon_{sat}>$ by simply using the $\varepsilon_{sat}$ at the average $<\rho>$. The overall density distribution for satellites is shown as the light purple curve in the bottom panel of Figure 1 and the average density is roughly $< \log(1+\delta) > \sim 1.0$, which corresponds to $\varepsilon_{sat} \sim 40\%$ in Figure 2.

We then use equation (10) to determine the mass function of all passive satellites using this average $<\varepsilon_{sat}> \sim 40\%$, plus the $\varepsilon_m$ from equation (8). If we wanted to determine, for example, the mass function of passive satellites in denser regions with $<\log(1+\delta)> \sim 2$, from Figure 2 we should use $\varepsilon_{sat} \sim 75\%$ in equation (10).

If both the $\varepsilon_m$ and $\varepsilon_{sat}$ had been determined solely from the colour data of the galaxies, then the agreement of the "predicted" red mass functions with the actual mass functions would not be surprising. However, as noted above, the form of $\varepsilon_m$ is a prediction from the $M^*$ and $\alpha_s$ of the *blue* population of galaxies (with no reference at all to the *red* population). In addition we have imposed the constraint that $\varepsilon_{sat}$ must be independent of mass from the separability of P10.

As we addressed in P10, we prefer the mass function as a better diagnostic tool to study galaxy evolution over the red fraction, since the red fraction reflects the *relative* $\phi$ of the red and blue galaxies at a given mass and contains no information on the shape of the mass functions with mass, i.e. $\phi(m)$. For instance, the red fraction alone doesn't require that the blue satellites are represented by a single Schechter function and that the red ones by a double Schechter function, as predicted from our model, or that both red and blue centrals have a single Schechter function. This is why the mass functions provide a powerful test of the model in both P10 and here.

The absolute normalization $\phi$ of these predictions in each panel is of course arbitrary, but the relative normalizations of the different components in each panel are fixed by equations (9-11). As shown in the two right panels in Fig. 13, the predictions from the model show excellent agreement with the data. For the centrals, the deviation from the model prediction at the highest masses is interpreted as the signature of modest post-quenching merging of centrals increasing the masses by about 25%. This deviation is due to that equations (9-11) are derived by neglecting the effect of merging.

For the star-forming galaxies that are centrals, and those that are satellites, a single Schechter function provides a fully satisfactory fit, over more than two orders of magnitude in stellar mass. It can be seen in Table 2 that the value of the $M^*$ of the star-forming centrals is identical to that of the star-forming satellites, i.e. within 0.02 ± 0.03 dex. This is further evidence for the "universality" of the $M^*$ of star-forming galaxies, complementing the demonstration in P10 that the $M^*$ are the same across a wide range of over-density $\delta$ and epoch.

Since $M^*$ of the star-forming mass function reflects the action of the mass-quenching process alone, this constancy of $M^*$ in different environments is a simple and direct consequence of the environment-independence of the mass-quenching process: the constant $\mu$ in our mass-quenching law is evidently invariant across all environments (and across all stellar masses and all epochs). The constancy of M* between star-forming centrals and satellites in Table 2 further illustrates the profound separability between mass and environment as drivers of galaxy evolution that we highlighted in P10.

For the passive centrals, we see a single Schechter function, as predicted, with a changed faint end slope $\Delta\alpha_s \sim 0.99 \pm 0.05$ relative to the star-forming centrals, as predicted. The $M^*$ is observed to be 0.09 ± 0.03 dex larger than for the star-forming galaxies. As noted above, this likely reflects some small degree of post-quenching merging of satellite galaxies into the centrals, boosting the masses of the latter by 25% on average. It is noticeable (and pleasing to the authors) that, we do *not* see the evidence for an associated change in $\alpha_s$ of the centrals (c.f. P10),

Table 2: Schechter function parameters for observed SDSS mass functions

| Sample | log (M*/M☉) | $\phi_1^*/10^{-3}$Mpc$^{-3}$ | $\alpha_1$ | $\phi_2^*/10^{-3}$Mpc$^{-3}$ | $\alpha_2$ |
|---|---|---|---|---|---|
| Blue central | 10.61±0.01 | ... | ... | 0.827±0.030 | -1.32±0.02 |
| Red central | 10.70±0.02 | 2.23±0.08 | -0.33±0.04 | ... | ... |
| Blue satellite | 10.59±0.02 | ... | ... | 0.196±0.013 | -1.56±0.03 |
| Red satellite | 10.61±0.02[a] | 0.935±0.07 | -0.49±0.08 | 0.130±0.042 | [-1.5][a] |

Notes to the Table:
(a) The M* for both Schechter components in the double Schechter function are assumed to be the same. The $\alpha_s$ of the second component is assumed to be -1.5.
(b) Following the convention in P10, we use subscript 1 in $\phi$ and $\alpha$ to refer to populations with $\alpha_s \sim$ -0.5 and subscript 2 to refer to populations with $\alpha_s \sim$ -1.5.

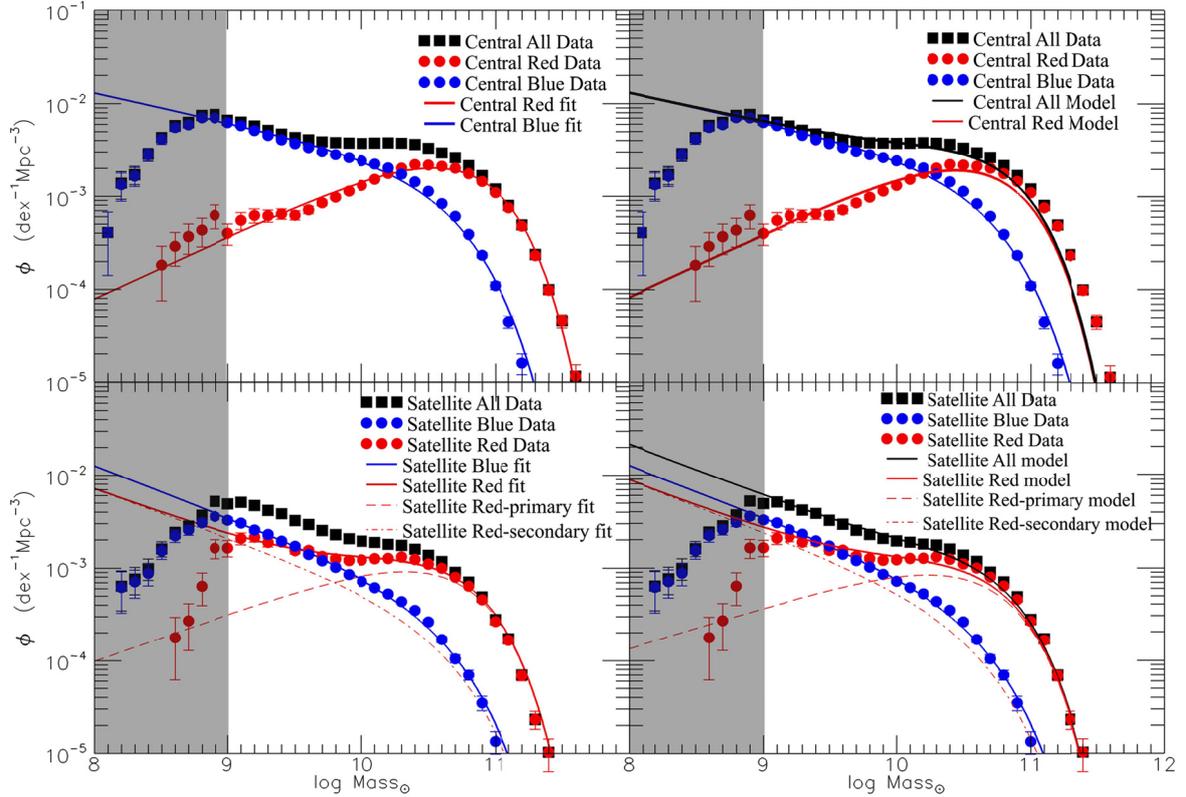

Figure 13. Galaxy Stellar Mass Function (GSMF) of SDSS galaxies as functions of blue and red galaxies (blue and red symbols respectively, with the sum in black) for centrals (upper two panels) and satellites (lower two panels). The two panels on the left show the observed data (points with statistical error bars) and the Schechter function fits as given in Table 2. The two panels on the right show the data and the model mass functions for the passive galaxies (and total population) that are computed based on the mass-function of active galaxies using equations (9) and (10), as described in the text. The grey shaded regions show where the data is seriously incomplete. For the centrals (upper right panel) there is a single passive component produced by mass-quenching. The deviation from the model prediction at the highest masses is interpreted as the signature of modest post-quenching merging of galaxies increasing the masses by about 25%. For the satellites (lower right panel) there are two components, one produced by mass-quenching and the other by satellite-quenching. For the passive satellites, there is little evidence for post-quenching merging.

in the sense that the $\Delta\alpha_s$ from the star-forming galaxies is "correct", with $\Delta\alpha_s \sim$ 1. As discussed at the end of the previous section, this is expected when we differentiate between centrals and satellites, as in this paper, since there is no change in the number density of the centrals. This change in $\alpha$ will be present when all passive galaxies in a given environment are lumped together, as was done in P10, because then there is a non-conservation of number.

Turning to the mass function of the passive satellites, this should be a double Schechter function, since these galaxies can have been quenched by either mass-quenching, or by satellite-quenching. The characteristic inflexion of the double Schechter function shape is indeed seen in the data in the Fig. 13. In Table 2, the parameters of the "primary" passive component are again exactly as expected relative to the shape of the star-forming satellite mass-function, with $\Delta\alpha_s$ = 1.07 ± 0.09 and $\Delta M^*$ = 0.02 ± 0.03 dex. We note as an aside, that this implies that the post-quenching merging of mass-quenched satellites must be completely negligible.

We can only estimate a ϕ* for the secondary (environment-quenched) passive component of the satellite population by assuming the same M* and $\alpha_s$ as for the star-forming galaxies (see above). When we do this, we find a ratio of $\phi^*_{red,sat,1}/\phi^*_{red,sat,2}$ = 7.2 ± 2.4, to be compared with the expectation derived above (immediately following equation 11) of 5.0. Looking at the ratio of the secondary passive component to the star-forming component of the satellites, we find $\phi^*_{red,sat,2}/\phi^*_{blue,sat}$ = 0.66 ± 0.22, to be compared with an expectation of 0.66 also from the same equation.

Finally we note that the ratio of the sum of the $\phi_2$* to the $\phi_1$* components, and thus the broad double Schechter form of the total mass function, is the same for centrals (0.37 ± 0.02) and satellites (0.35 ± 0.05). This is a reflection of an "inevitability" of the overall double Schechter mass function which we explore further in a later paper.

Looking back to Figures 9 & 10, it is clear that the inter-relationships between the Schechter parameters of the mass functions of star-forming and passive satellite galaxies shown quantitatively here are also maintained in individual bins of halo mass.

In the P10 formalism, the faint end slope $\alpha_s$ of the blue population is one of the few input parameters in the model. It is noticeable in Table 2 that this parameter is rather different for the centrals and satellites, by 0.24 ± 0.04, even though both sets of galaxies exhibit the correct $\Delta\alpha_s \sim 1$ between the passive and star-forming populations. This difference in $\alpha_s$ for the star-forming galaxies is a manifestation of the fact that the fraction of galaxies that are satellites is not completely independent of stellar mass, but increases slowly to lower masses. This requires the small difference in $\alpha_s$ between centrals and satellites that we see here.

A steeper $\alpha_s$ of the mass function for *all* star-forming satellites can be also expected from Figure 9. It is clear from Figure 9 that the mass function of the star-forming satellites, in in a given parent halo mass, has an $\alpha_s \sim -1.4$, which is the same as the $\alpha_s$ of the star-forming mass function in the most under dense D1 density quartile (P10) and is also similar to the $\alpha_s$ of the star-forming centrals. However, due to the limit on the stellar mass of the most massive satellite that a given host halo can support because of the mass of the central (given by the diagonal light blue lines in the two bottom panels in Figure 8), the mass functions of the star-forming satellites are truncated at a progressively higher stellar mass with increasing halo mass. Therefore, when we add together the mass functions of the star-forming satellites over a range of halo masses, we will get a steeper $\alpha_s$ of the overall composite mass function simply because of the range of truncation masses. Since the faint end slope of the mass function of star-forming galaxies is one of the few inputs to our formalism, this effect is immaterial to our results. What is relevant is the inter-relationships between the parameters for different components of the mass function, and these are as predicted.

The existence of the second component of passive galaxies, i.e. the upturn in the mass function at low masses, is also clearly seen in the deep photo-*z* based COSMOS analysis of Drory et al. (2009) and Ilbert et al. (2010) at z < 0.6; in the GAMA survey of Baldry et al. (2012) at z < 0.06. This becomes more challenging at higher redshifts since the complete mass of the passive galaxies is increasing with redshifts, plus that the fact that the environment-quenching process is on average weaker at higher redshifts. Thus one needs to go even deeper to observe the second component of passive galaxies at higher redshifts.

## 7. SUMMARY

In an earlier paper (P10) we developed a new approach to the study of the evolving galaxy population in SDSS and zCOSMOS. This was based on the identification of the underlying simplicities of the galaxy population, as characterized by the stellar masses, star-formation rates and Mpc-scale environments of galaxies. In that paper, the environments were described using a 5th nearest neighbor overdensity δ.

This lead to the identification of two distinct processes that quench star-formation in galaxies: mass-quenching which is evidently independent of environment, and environment-quenching which must be independent of stellar mass. As discussed in P10, these two quenching processes, coupled with the global cosmic evolution in the specific star-formation rate (sSFR) largely control the growth in stellar mass of galaxies, at least since $z \sim 2$, with only a relatively modest contribution from merging in the denser environments.

One of the unexpected successes of this approach was the prediction of the precise Schechter function form for the mass functions of the active and passive galaxies and the accurate reproduction of the inter-relationships between the Schechter parameters M*, α and ϕ* for these two populations and in high and low density environments.

In P10, we speculated that our environmental-quenching process could well be linked to satellite galaxies. This speculation was based on the fact that our environment quenching efficiency $\varepsilon_\rho(\rho,m,z)$ shared several key properties with $f_{sat}(\rho,m,z)$, the fraction of galaxies that are satellites (as opposed to centrals) in semi-analytic models of the galaxy population (e.g. Kitzbichler & White 2007). Both $\varepsilon_\rho$ and $f_{sat}$ are observed to depend on over-density δ, but not on either stellar mass *m* or redshift *z* (at least out to $z \sim 1$).

In the current paper, we have been able to check and confirm this speculation and to significantly extend the P10 approach, by studying the central and satellite galaxy dichotomy in the SDSS sample at low redshift, using the Yang et al group catalogue.

We find the following:

1. Central galaxies have a red fraction $f_{red}$ that is largely independent of over-density δ but depends on stellar mass. The dependence of $f_{red}$ on stellar mass for centrals is exactly the same as that of the entire galaxy population in the lowest density environments in our earlier analysis. Satellite galaxies in contrast have an $f_{red}$ that increases monotonically with δ, as well as stellar mass, and is always larger than that of the centrals.

2. A satellite-quenching efficiency parameter $\varepsilon_{sat}$, defined to be analogous to our earlier mass- and environment-quenching efficiency parameters and to represent the fraction of star-forming (previously central) galaxies that are quenched when they become satellites of another galaxy, is found to be independent of stellar mass, in agreement with van den Bosch et al. (2008), P10 and Quadri et al. (2011) for *z* = 2. However,

$\epsilon_{sat}$ is found to strongly depend on environment.

3. The fraction of satellite galaxies that are passive at a given stellar mass correlates better with a nearest-neighbor over-density parameter $\delta$. At these scales, $\delta$ is tracing primarily the location within the group/halo rather than the surrounding large-scale density field exterior to the group. The richness can be taken as a proxy for the halo mass. This suggests that our environment-quenching process is driven by over-density (location within a halo) and not by the halo dark matter mass, or of course by the stellar mass.

4. The star-formation rates of star-forming satellites are also independent of the halo mass and are essentially identical to those of star-forming central galaxies.

5. The fact that the M* of satellites is the same over a wide range of at least two orders of magnitude in halo mass above $10^{12}M_\odot$ indicates the mass-quenching process for satellites operates independently of halo mass.

6. The tight correlation between the stellar mass and halo mass for star-forming central galaxies, makes it difficult to distinguish between a picture involving a link between the quenching rate and the SFR (as in P10) (which is equivalent to a limit to the stellar mass attained by a galaxy) or a limit to the mass of dark matter halo that can support star-formation. However, the fact that M* for star-forming centrals and for satellites is essentially identical suggests that mass-quenching operates in the same way on both centrals and satellites. With some caveats, this argues against links between mass-quenching and dark matter halo mass for all galaxies. The caveats are that the mass-quenching of satellites could be driven by their own sub-haloes, or that they have increased their stellar masses very little since they have become satellites, so that their distribution of masses reflects processes operating when they were actually centrals, rather than when they were satellites.

7. The four mass functions representing centrals and satellites, split into star-forming and passive galaxies, follow the precise quantitative relations expected for our simple model in which centrals are only quenched through (environment-independent) mass-quenching, but the satellites are quenched through both mass-quenching and the mass-independent environment-quenching process identified in P10.

8. The effects of post-quenching merging, which modify slightly these relationships, are shown to occur only in central galaxies. The effects of post-quenching merging are subtly different from those in P10, producing a change in $M^*$ without an associated change in $\alpha_s$, reflecting the fact that it is satellites that are merging with centrals. We estimate that typical central galaxies can have accreted only 25% of their mass through merging after being quenched.

These new findings significantly extend our earlier analysis and formalism that we presented in P10. They show that all of the environmental influences on the quenching of galaxies identified in P10 are driven by satellite galaxies.

This environmental quenching of satellites does not apparently depend on the stellar mass of the satellite, nor to a large degree, does it appear to depend on the mass of the parent halo. Rather, it does depend on local over-density, i.e. the position within the halo. The mass-quenching of galaxies, which we showed in P10 dominated the overall evolution of the galaxy population, also appears to know little about the mass of the halo since the effects of mass-quenching on satellites do not reflect the (present-day) mass of their parent haloes and because mass-quenching appears to operate in the same way on centrals and satellites.

## ACKNOWLEDGEMENTS

We are very grateful to Xiaohu Yang and his collaborators for generously providing the DR7 version of their group catalogue. Since the publication of our original paper P10 we have had stimulating discussions with many colleagues, too numerous to mention by name but thanked nonetheless. We thank the referee for useful comments, which helped us to identify parts of the manuscript that required explanation. We acknowledge NASA's IDL Astronomy Users Library, the IDL code base maintained by D. Schlegel, and the *kcorrect* software package of M. Blanton. This research has been supported by the Swiss National Science Foundation.

**APPENDIX A**

Starting from the standard continuity equation, for the rate of change in number density of star-forming galaxies $\phi_{blue}$ per unit logarithmic mass bin, at fixed mass and environment,

$$\frac{\partial \phi_{blue}(t)}{\partial t} + \frac{\partial}{\partial \log m} \cdot [\phi_{blue}(t) \frac{\partial \log m}{\partial t}] = -[\lambda_m(t) + \kappa_-(t)]\phi_{blue}(t)$$

(A1)

where $\lambda_m$ is the mass-quenching rate which we would like to derive.

We assume, as in P10, that the merging term $\kappa_-$ is either independent of stellar mass or that it is negligible in under-dense regions. The second assumption at least seems highly plausible, and the first is not unreasonable.

The second term in the left hand side of (A1) can be reformed as

$$\frac{1}{\phi_{blue}(t)}\frac{\partial}{\partial \log m} \cdot [\phi_{blue}(t)\frac{\partial \log m}{\partial t}]$$
$$= \frac{1}{\phi_{blue}(t)}\frac{\partial \phi_{blue}(t)}{\partial \log m}\frac{\partial \log m}{\partial t} + \frac{\partial}{\partial \log m}[\frac{\partial \log m}{\partial t}]$$
$$= sSFR(t)[\frac{\partial \log \phi_{blue}(t)}{\partial \log m} + \frac{\partial \log sSFR(t)}{\partial \log m}] \quad (A2)$$
$$= sSFR(t)(\alpha + \beta)$$

where we have used the definition of sSFR:

$$sSFR(t) = \frac{SFR(t)}{m} = \frac{d \log m}{dt}\ln 10 \quad (A3)$$

The $\alpha$ term in (A2) is the logarithmic slope of the mass function and $\beta$ is logarithmic slope of the sSFR-mass relation, defined as:

$$\alpha = \frac{\partial \log \phi_{blue}(t)}{\partial \log m} = (1+\alpha_s) - \frac{m}{M^*} \quad (A4)$$

$$\beta = \frac{\partial \log sSFR(t)}{\partial \log m} \quad (A5)$$

Inserting equation (A2) into (A1), we recover the middle equation in the set of equations given as Equation (10) in P10.

$$\frac{1}{\phi_{blue}(t)}\frac{\partial \phi_{blue}(t)}{\partial t} = -sSFR(t)(\alpha+\beta) - \lambda_m(t) - \kappa_-(t)$$
$$= -sSFR(t)(1+\alpha_s+\beta) + \frac{SFR(t)}{M^*} - \lambda_m(t) - \kappa_-(t)$$

(A6)

It should be kept in mind that the above derivations are done at fixed mass and environment and we have dropped the $|_{m,\rho}$ notation in all parameters in all equations in this section for clarity.

To keep a constant shape (in terms of $\alpha_s$ and M*) of the star-forming mass function with time, a key observationally motivated axiom of our analysis, clearly requires $d\log\phi_{blue}/dt$ to be independent of mass. This means that the left hand side of (A6) must be independent of mass.

Turning to the right hand side of (A6), since we have assumed that the merging term $\kappa_-$ is mass independent (or negligible) and because we also assume $\beta=0$, i.e. the sSFR is mass independent, then the only solution of $\lambda_m$ to (A6) is

$$\lambda_m(t) = SFR(t)/M^* + C(t) \tag{A7}$$

where $C(t)$ is some mass independent, but possibly time-dependent term. Given the fact that low mass galaxies (i.e. $\varepsilon_m \sim 0$) in under-dense regions (i.e. $\varepsilon_\rho \sim 0$) are almost all star-forming galaxies at $z \sim 0$ (see Figure 6 in P10) and at higher redshifts at least up to $z \sim 1$ (see Figure 8 and equation 6 in P10), this requires that any quenching rate $C(t)$ in (A7) must be negligible at low masses, and thus at all masses. Therefore, we have

$$\lambda_m(t) = SFR(t)/M^* \tag{A8}$$

## APPENDIX B

Following the continuity equations of eq.(10) in P10, when the M* of the mass function of the star-forming galaxies is established (i.e. M* will keep constant afterwards), with mass-quenching only (i.e. for all galaxies in the most under dense regions or for centrals in all environments), the mass function of the star-forming galaxies $\phi_{blue}(t)$ at given $m$ is given by:

$$\phi_{blue}(t) = \phi_{blue}(t_0) e^{\int_{t_0}^{t} -(1+\alpha_s+\beta) sSFR(t') \, dt'} \tag{B1}$$

Again, all the derivations in this section are done at fixed mass and we have dropped the $|_m$ notation in all parameters in all equations for clarity.

In (B1) $\phi_{blue}(t_0)$ is the mass function of the star-forming galaxies at an earlier time $t_0$.

The change of the mass function of the mass-quenched passive galaxies $\phi_{red}$ at given $m$ is given by:

$$\frac{d\phi_{red}(t)}{dt} = \phi_{blue}(t) \lambda_m = \phi_{blue}(t) \frac{SFR(t)}{M^*} \tag{B2}$$

Inserting (B1) into (B2), it follows that

$$\phi_{red}(t) = \phi_{blue}(t_0) \int_{t_0}^{t} \frac{SFR(t')}{M^*} e^{\int_{t_0}^{t'} -(1+\alpha_s+\beta) sSFR(t'') \, dt''} dt'$$

$$= \phi_{blue}(t_0) \int_{t_0}^{t} m \frac{sSFR(t')}{M^*} e^{\int_{t_0}^{t'} -(1+\alpha_s+\beta) sSFR(t'') \, dt''} dt' \tag{B3}$$

Since we calculate the above integral at fixed $m$, we can move $m$ out of the integral in the second equation of (B3):

$$\phi_{red}(t) = \phi_{blue}(t_0) \frac{1}{-(1+\alpha_s+\beta)} \frac{m}{M^*} \cdot$$

$$\int_{t_0}^{t} -(1+\alpha_s+\beta) sSFR(t') e^{\int_{t_0}^{t'} -(1+\alpha_s+\beta) sSFR(t'') \, dt''} dt'$$

$$= \phi_{blue}(t_0) \frac{1}{-(1+\alpha_s+\beta)} \frac{m}{M^*} e^{\int_{t_0}^{t} -(1+\alpha_s+\beta) sSFR(t') \, dt'}$$

$$= \phi_{blue}(t) \frac{1}{-(1+\alpha_s+\beta)} \frac{m}{M^*} \tag{B4}$$

Putting equation (B2) and (B4) together, gives

$$\frac{1}{\phi_{red}(t)} \frac{d\phi_{red}(t)}{dt} = -sSFR(t)(1+\alpha_s+\beta)$$

$$= \frac{1}{\phi_{blue}(t)} \frac{d\phi_{blue}(t)}{dt} \tag{B5}$$

$$\frac{d\ln\phi_{red}(t)}{dt} = \frac{d\ln\phi_{blue}(t)}{dt}$$

From (B4), it's straightforward to show that

$$f_{red}(m, \rho_0) = \varepsilon_m(m) = \frac{\phi_{red}(t)}{\phi_{red}(t) + \phi_{blue}(t)}$$

$$= \frac{1}{1 - (1+\alpha_s+\beta) \frac{M^*}{m}} \tag{B6}$$

which is the equation (8) as derived in section 4.3.

It is clear from equation (B4) that the mass function of the mass-quenched passive galaxies is also a Schechter function with exactly the same $M^*$ but with a modified faint-end slope with $\Delta\alpha_s = 1$. Because M* and $\alpha_s$ for star-forming galaxies are observed to be invariant with time since early epochs, it follows that the integrated $\phi(m)$ of the mass-quenched passive galaxies will also build up over time with the same constant M* and modified $\alpha_s$, as required by equation (B4).

Equation (B5) further makes it clear that the increase in $\log \phi$ of the passive galaxies over time matches exactly that of the star-forming galaxies. This also explains why in this case we have a steady state color distribution, i.e. that the $\varepsilon_m$ is constant over cosmic time and $df_{red} / dt = 0$ at given $m$, as discussed in section 4.3.

The two components, with the same M*, with $\alpha_s$ differing by a constant amount and with their two $\phi^*$ increasing in step, thus produce a characteristic double Schechter function shape (for the overall population of galaxies) that, once established, should not change.

Any mass independent environment quenching will not change M* and $\alpha_s$. Thus it will not change the shape of the mass functions of star-forming and passive galaxies. The environment quenching will act to decrease of the normalization $\phi$ of the star-forming mass function and increase of the $\phi$ of the component of the passive mass function that has the same $\alpha_s$. Thus while it controls the relative normalization of star-forming and passive galaxies, and thus the red fraction, it has no effect on the shape of the combined mass-function. We will further demonstrate this in detail in a future paper.